\documentclass[twocolumn,  twocolappendix, times]{aastex701}

\usepackage[T1]{fontenc}
 \hypersetup{linkcolor=magenta, 
 citecolor=blue, 
 filecolor=green, 
 urlcolor=magenta}

\usepackage{enumitem}
\usepackage[separate-uncertainty=true]{siunitx} 
\usepackage{float}
\DeclareSIUnit \parsec {pc}
\DeclareSIUnit \year {yr}
\DeclareSIUnit \erg {erg}
\DeclareSIUnit \day {day}

\begin{document}

   \title{Constraints on the VHE counterpart of two binary black hole mergers observed by the MAGIC and CTAO LST-1 telescopes}

\footnotesize
\author[]{K.~Abe}\email{}\affiliation{Department of Physics, Tokai University, 4-1-1, Kita-Kaname, Hiratsuka, Kanagawa 259-1292, Japan}
\author[]{S.~Abe}\email{}\affiliation{Institute for Cosmic Ray Research, University of Tokyo, 5-1-5, Kashiwa-no-ha, Kashiwa, Chiba 277-8582, Japan}
\author[]{J.~Abhir}\email{}\affiliation{ETH Zürich, CH-8093 Zürich, Switzerland}
\author[]{A.~Abhishek}\email{}\affiliation{INFN and Universit`a degli Studi di Siena, Dipartimento di Scienze Fisiche, della Terra e dell'Ambiente (DSFTA), Sezione di Fisica, Via Roma 56, 53100 Siena, Italy}
\author[]{V.~A.~Acciari}\email{}\affiliation{Institut de Fisica d'Altes Energies (IFAE), The Barcelona Institute of Science and Technology, Campus UAB, 08193 Bellaterra (Barcelona), Spain}
\author[]{F.~Acero}
\email{}\affiliation{Universit\'e Paris-Saclay, Universit\'e Paris Cit\'e, CEA, CNRS, AIM, F-91191 Gif-sur-Yvette Cedex, France}
\email{}\affiliation{FSLAC IRL 2009, CNRS/IAC, La Laguna, Tenerife, Spain}
\author[]{A.~Aguasca-Cabot}\email{}\affiliation{Departament de Física Quàntica i Astrofísica, Institut de Ciències del Cosmos, Universitat de Barcelona, IEEC-UB, Martí i Franquès, 1, 08028, Barcelona, Spain}
\author[]{I.~Agudo}\email{}\affiliation{Instituto de Astrofísica de Andalucía-CSIC, Glorieta de la Astronomía s/n, 18008, Granada, Spain}
\author[]{I. Albanese}\email{}\affiliation{INFN Sezione di Padova and Universit\`a degli Studi di Padova, Via Marzolo 8, 35131 Padova, Italy}
\author[]{D.~Ambrosino}
\email{}\affiliation{INFN Sezione di Napoli, Via Cintia, ed. G, 80126 Napoli, Italy}
% ...

\collaboration{all}{(the MAGIC and CTAO-LST Collaborations)}
\correspondingauthor{(alphabetical order) J. Jiménez Quiles, M. Pecimotika, M. Seglar Arroyo, A. Simongini.}

\begin{abstract}

    We present very-high-energy gamma-ray observations of two binary black hole merger candidates, GW240615\_113620 and GW241125\_010116, performed with the Major Atmospheric Gamma Imaging Cherenkov (MAGIC) telescopes and the first Large-Sized Telescope of the Cherenkov Telescope Array Observatory's (CTAO LST-1). 
    GW240615\_113620 was the best localized event of the fourth observing run of the LIGO-Virgo-KAGRA gravitational waves interferometers.
    GW241125\_010116 was temporally and spatially coincident with a sub-threshold short-duration burst detected with the \textit{Swift}-Burst Alert Telescope (BAT), the \textit{Swift}-X-Ray Telescope (XRT) and the \textit{Einstein Probe} Follow-up X-ray Telescope (FXT).
    We observed the two events in stereoscopic mode, taking advantage of the improved sensitivity of joint MAGIC+LST-1 observations.
    No detection was achieved in the GeV-TeV gamma-ray band for any of the two sources. 
    The unfavourable observing conditions of both events posed a challenge for a standard analysis and therefore a non standard analysis was necessary for both objects. 
    Owing to the small localization area and the association with a GRB-like burst respectively, these events represented an unprecedented opportunity to study in details the electromagnetic emission from binary black holes merger events and, in particular, we discussed two theoretical models that predict a detectable gamma-ray emission and the possible future applications. 

\end{abstract}


\section{Introduction}
\setcounter{footnote}{0}

    The first detection of a gravitational wave (GW) from a binary black hole (BBH) system merger was a historical breakthrough, marking the beginning of multimessenger astronomy as we know it today \citep{abbott2016observation}.
    This led to a technological advance on GW interferometers, currently the Advanced Laser Interferometer Gravitational Wave Observatory (LIGO; \citealt{aasi2015advanced}), Advanced Virgo+ \citep{acernese2014advanced,flaminio2020status} and Kamioka Gravitational Wave Detector (KAGRA; \citealt{aso2013interferometer}), and corresponding LIGO-Virgo-KAGRA (LVK) Collaboration(s). 
    Improvements in broadband sensitivity led to growth in the number of candidate GW events detected throughout the first four observing runs, achieving a total of 390 confirmed events so far\footnote{\url{https://gwosc.org/eventapi/html/GWTC/}}. 
    Among them, 11 were observed during the first two observing runs (O1 and O2; \citealt{abbott2019gwtc}), 79 during the first and second half of the third run (O3a and O3b; \citealt{abbott2021gwtc, abbott2023gwtc, abbott2024gwtc}), 139 during the first part of the fourth run (O4a; \citealt{2024ApJ...970L..34A, 2025arXiv250708219T}), and the remaining 161 during the second part of the fourth run (04b; \citealt{gwtc5a, gwtc5b}).  
    The vast majority of confirmed detections originate from BBH systems, with total masses in the range $3.4 - 236 M_\odot$, corresponding to chirp masses of $1.2 - 101 M_\odot$ and distances up to 8.3 Gpc. 
    Among the four observing runs, only two events were classified as binary neutron star (BNS; \citealt{abbott2017gw170817, abbott2020gw190425}) and three neutron-star–black-hole binaries (NSBHs; \citealt{abbott2021observation, 2024ApJ...970L..34A}). 

    Detecting electromagnetic (EM) counterparts of GW events is one of the main challenges of current astrophysics and many facilities around the world and in space are dedicating their efforts to make this detection possible. 
    The historical detection of GW170817 from a BNS merger by LIGO and Virgo \citep{abbott2017gw170817, abbott2019properties, abbott2019tests} demonstrated the importance of multimessenger astronomy (the first time being the electromagnetic and neutrino detection of SN~1987A; e.g. \citealt{hirata1988observation}). 
    Remarkably, the GW was spatially and timely coincident with a short gamma-ray burst (GRB), GRB~170817A, detected by the \textit{Fermi} Gamma-Ray Burst Monitor (GBM; \citealt{goldstein2017ordinary}) and the INTErnational Gamma-Ray Astrophysics Laboratory (\textit{INTEGRAL}; \citealt{savchenko2017integral}) space satellites, providing strong evidence of the association between BNS mergers and GRBs \citep{abbott2017gravitational}. 
    Soon after the initial trigger, an extensive follow-up campaign was launched across the entire EM spectrum leading to the detection of the optical/NIR counterpart AT~2017gfo, followed by X-rays and radio detection of the GRB afterglow (\citealt{2017ApJ...848L..12A} and references therein). 
    The afterglow was extensively followed in the gamma-ray band, spanning over 10 orders of magnitude in energy by space and ground-based telescopes. However, no significant detection was achieved in any gamma-ray band \citep{goldstein2017ordinary, 2017GCN.21534....1K, martinez2017ligo, nakahira2017ligo, savchenko2017integral, 2017GCN.21746....1S, verrecchia2017agile, li2018insight}. 
    The High Energy Stereoscopic System (H.E.S.S.) array of imaging atmospheric Cherenkov telescopes (IACTs) conducted deep observations in the very-high-energy (VHE; $E >$ \SI{50}{\giga\electronvolt}) gamma-ray regime, achieving upper limits (U.L.s) in the range of \num{0.13} -- \SI{23.7}{\tera\electronvolt} \citep{abdalla2017tev}. 
    Observations were performed with two different cadences: rapid observations from 0.22 to 5.2 days post-merger, and late-time observations 200 days post-merger \citep{abdalla2020probing}. 
    The Major Atmospheric Gamma Imaging Cherenkov (MAGIC) telescopes observed the event between 150 and 300 days after the trigger, achieving a flux U.L. of $3.6\times 10^{-12}\,\text{erg}\,\text{cm}^{-2}\,\text{s}^{-1}$ at energies $>$ \SI{0.4}{\tera\electronvolt} \citep{stamerra2022follow}. 
    Notably, the properties of the event, with low interstellar medium density and relatively large viewing angle, were not favourable for VHE detection. 
    However, the detectability of a TeV component from a GW counterpart was hinted by the observation of the short GRB~160821B by MAGIC, although only a 3$\sigma$ excess was found \citep{acciari2021magic}. 

    Owing to the extraordinary event of 2017, the existence of an EM counterpart of a GW from a BNS merger is no longer a question. 
    On the other hand, no EM counterpart of BBH events has still been observationally confirmed, despite the high number of GW detected (of the order of 100 times greater than BNS events) and different proposed candidates (e.g. \citealt{connaughton2016fermi, connaughton2018interpretation, graham2020candidate}). 
    A significant EM emission from stellar-origin BBH mergers (i.e. both masses of the order of 10 -- 100 M$_\odot$) requires particular physical conditions that are still not fully comprehended regarding binary formation channels and the interaction with their environment, nor observationally constrained. 
    One of the most promising channels to produce EM emission is the interaction of the binary or the merger remnant with the gaseous accretion disk of an active galactic nucleus (AGN; e.g. \citealt{bartos2017rapid, stone2017assisted, mckernan2019ram}). 
    A large number of stellar-mass black holes (BHs) is expected to populate the environment of an AGN as a consequence of dynamical friction-driven migration \citep{1993ApJ...408..496M, bartos2013g2, bartos2017rapid}, with both theoretical arguments and observational evidences indicating a number of $2\times10^4$ BHs within the central parsec \citep{bahcall1976star, hailey2018density}.
    In this crowded environment, BBH systems can form via dynamical encounters of individual BHs \citep{o2009gravitational}. 
    If a BBH is embedded in the dense and gaseous accretion disk of an AGN, the gravitational merger is accelerated ($\le$ \SI{1}{\mega\year}) and the amplified accretion within the disk can give rise to EM emission \citep{bartos2017rapid,xue2025determines}. 
    In this paper, we focus exclusively on the EM emission of BBH mergers embedded in an AGN disk, which should account for $25-80\%$ of all BBH mergers \citep{ford2022binary}.
    Such emission involves X-rays and soft gamma-rays produced by synchrotron radiation from the relativistic jets that are launched after the merger or during the accretion \citep{bartos2013gravitational, bartos2017rapid}. 
    In addition, \cite{kimura2021outflow} studied the case of kicked BBH mergers, where they identified that at most \SI{10}{\percent} would have enough kick velocity and kick direction to re-enter the AGN disk and produce EM emission.
    On the other hand, the synchrotron emission alone cannot produce photons up to tens of GeV, leading to the necessity of additional radiative processes such as inverse Compton scattering. 
    Notably, previous studies on BBH merger counterparts by H.E.S.S. \citep{abdalla2021hess} and VERITAS \citep{2019ICRC...36..782S} primarily focused on the GRB-like scenario without discussing potential origins of VHE gamma-ray emission.
    
    We report the observations and analysis of two BBH candidates detected by LVK during the O4b run: GW240615\_113620 (hereafter, GW240615) and GW241125\_010116 (hereafter, GW241125). 
    We followed the two events with the MAGIC telescopes and the Cherenkov Telescope Array Observatory's (CTAO's) first Large-Sized Telescope (LST-1) in stereo-joint configuration, $\sim$ 14 hours and $\sim$ 19 hours after the initial trigger, respectively. 
    In both cases, the very good localization of those sources enabled their treatment as point-like sources for observations by IACTs.  

    This paper is structured as follows: we present the GW data in Sect.~\ref{Sec:2} and the observation and analysis of the two events in Sect.~\ref{Sec:3}, ~\ref{Sec:4} and ~\ref{Sec:5}. 
    Our results are discussed in Sect.~\ref{Sec:6}. 
    Theoretical models to contextualise our upper limits are discussed in Sect. ~\ref{sec:agns}. 
    We draw the final conclusions in Sect.~\ref{Sec:8}.

\section{The two BBH events}\label{Sec:2}
\subsection{GW240615} 

    GW240615 (S240615dg) was identified by the LVK interferometers on the 15th of June 2024 at 11:36:20.724 UTC.
    This event was immediately classified as a BBH merger with a probability of \SI{99}{\percent} and a false alarm rate (FAR) of \SI{3.2e-10}{\hertz} (FAR = 1 per 100.04 years). 
    The luminosity distance, marginalized over the whole sky, is $D_{\rm L} = 1560^{+290}_{-450}\,\rm Mpc$. 
    GW240615 stands as the best localized event of all LVK observing runs, with a \SI{90}{\percent} confidence area of only \SI{5}{\square\deg} \citep{2024GCN.36669....1L, 2024GCN.36704....1L, gwtc5a}. 
    This event was associated to the merger of two BHs with a total mass of $M_{\rm tot} = 60.4^{+3.5}_{-2.5}\, M_\odot$ (of which $M_1 = 34.0^{+5.4}_{-3.7}\, M_\odot$ and $M_2 = 26.4^{+4.0}_{-4.8}\, M_\odot$), with an equivalent chirp mass of $\mathcal{M}_{\rm c}= 25.9^{+1.7}_{-1.2}\, M_\odot$.
    The reported network signal-to-noise ratio is 26.4 \citep{gwtc5a, gwtc5b}. 

    Taking advantage of the small confidence area, the event was followed up by optical ground-based telescopes \citep{2024GCN.36705....1A, 2024GCN.36698....1P, 2024GCN.36972....1R} and the \textit{Swift} space telescope \citep{2024GCN.36837....1E, 2024GCN.36687....1R}, although no significant EM counterpart was found in any band.
    More details about this event can be found in the official GraceDB page\footnote{\url{https://gracedb.ligo.org/superevents/S240615dg}}.

\subsection{GW241125}

    GW241125 (S241125n) was identified by the LVK interferometers on the 25th of November 2024 at 01:01:16.780 UTC with a \SI{90}{\percent} confidence area of \SI{76}{\square\deg} \citep{2024GCN.38305....1L}. 
    This event was immediately classified as a BBH merger with a probability of \SI{99}{\percent} and FAR of \SI{9.5e-10}{\hertz} (FAR = 1 per 33.347 years). 
    The best estimate of the luminosity distance of the event is $D_{\rm L} = 4700^{+3900}_{-2200}\,\rm Mpc$ \citep{gwtc5a}. 
    This event was associated to the merger of two BHs with a total mass of $M_{\rm tot} = 106^{+23}_{-22}\, M_\odot$ (of which $M_1 = 60.0^{+14}_{-13}\, M_\odot$ and $M_2 = 47^{+14}_{-17}\, M_\odot$), with an equivalent chirp mass of $\mathcal{M}_{\rm c}= 45^{+10}_{-10}\, M_\odot$.
    The reported network signal-to-noise ratio is 11 \citep{gwtc5a, gwtc5b}. 
    More details about this event can be found in the official GraceDB page\footnote{\url{https://gracedb.ligo.org/superevents/S241125n/}}.  

    Despite having a very high uncertainty on the distance and localization, this GW represents a good target for EM searches because it originated from BHs in the pair-instability mass gap (if we consider a lower limit of $M_{\rm BH} \simeq 45 M_\odot$; \citealt{gerosa2021hierarchical}). 
    BHs of this kind challenge the current understanding of stellar evolution, and are predicted to be the remnant of a previous gravitational merger \citep{abbott2021gwtc, gerosa2021hierarchical, woosley2021pair}, increasing the probability of this BBH to have happened in an AGN disk \citep{tagawa2020formation, yang2019hierarchical, gerosa2021hierarchical}.

    At the time of the initial detection, GW241125 gained a lot of interest in the community thanks to the putative identification of a sub-threshold EM counterpart by the \textit{Swift} Burst Alert Telescope (BAT; \citealt{2024GCN.38305....1L, 2024GCN.38313....1L, 2024GCN.38308....1D}). 
    The burst did not pass the automatic repointing criteria of \textit{Swift}; however, as described in \protect\citet{2024GCN.38308....1D}, the telescope was triggered by the Gamma-ray Urgent Archiver for Novel Opportunities (GUANO).
    Spectral analysis of the putative counterpart revealed a Comptonized spectrum peaking at $E =$ \SI{49}{\kilo\electronvolt} with photon index $-1.2$ and a flux in the range between 15 -- \SI{350}{\kilo\electronvolt} of $1.1^{+0.2 }_{-0.3}\times10^{-7}\,\rm erg\,cm^{-2}\,s^{-1}$ \citep{2024GCN.38308....1D, 2024GCN.38351....1D}. 
    Following the initial trigger from \textit{Swift}, extensive follow-ups by many facilities were performed, however no significant EM counterpart was identified in any band. 
    The \textit{INTEGRAL} telescope reported an upper-limit of \SI{1.4e-7}{\erg\per\square\centi\meter\per\second} in the 75 -- \SI{2000}{\kilo\electronvolt} energy band assuming a short GRB spectrum \citep{2024GCN.38311....1S}. 
    The \textit{Fermi}-GBM obtained an upper-limit of \SI{5.5e-7}{\erg\per\square\centi\meter\per\second} in the 75 -- \SI{2000}{\kilo\electronvolt} energy band \citep{2024GCN.38316....1S} assuming a hard spectrum, while \textit{Konus}-Wind yielded an upper-limit of \SI{2.3e-7}{\erg\per\square\centi\meter\per\second} in the 10 -- \SI{1000}{\kilo\electronvolt} energy band \citep{2024GCN.38321....1R}. 
    A possible X-ray counterpart was detected by both modules of the \textit{Einstein Probe} Follow-up X-ray Telescope (EP-FXT) at \SI{94}{\kilo\second} after the initial trigger. 
    The position of the X-rays candidate counterpart falls within the 5-arcmin BAT error circle. 
    The derived average unabsorbed flux is \SI{1.17(1.18:0.63)e-13}{\erg\per\square\centi\meter\per\second} in the 0.5 -- \SI{10}{\kilo\electronvolt} band \citep{2024GCN.38345....1W}. 
    The position of the EP-FXT counterpart is consistent with an uncatalogued X-ray source detected by \textit{Swift} X-ray telescope (XRT) between \SI{55}{\kilo\second} and \SI{74}{\kilo\second} after trigger, for which a peak flux of \SI{9(4)e-14}{\erg\per\square\centi\meter\per\second} was derived in the 0.2 -- \SI{10}{\kilo\electronvolt} band \citep{2024GCN.38324....1P}. 
    The afterglow was extensively followed by optical telescopes around the world, although no detection was achieved \citep{2024GCN.38396....1A, 2024GCN.38329....1B, 2024GCN.38314....1C, 2024GCN.38328....1J, 2024GCN.38350....1K, 2024GCN.38325....1M, 2024GCN.38333....1R, 2024GCN.38322....1S, 2024GCN.38317....1W}.

\section{The Joint VHE Observation Campaigns}\label{Sec:3}

    The LST-1 is the first telescope of the 23-meter diameter telescopes array, designed to be part of the forthcoming CTAO in the Northern Hemisphere (CTAO-N; \citealt{cta2018science}). 
    Located at the Observatorio del Roque de los Muchachos in La Palma (Canary Islands, Spain). LST-1 features a reflective surface of approximately \SI{400}{\square\meter} and a camera equipped with high quantum efficiency photomultiplier tubes. 
    The telescope has a field of view (FoV) of \SI{4.3}{\deg}. 
    These characteristics make it particularly well-suited for detecting VHE gamma rays at the lower end of the spectrum, starting from a few tens of \unit{\giga\electronvolt} \citep{Abe2023}. 

    The MAGIC telescopes are two 17-meter diameter IACTs located at the same observatory as the LST-1, sensitive to VHE gamma rays from around $\sim$ \SI{50}{\giga\electronvolt} up to tens of \unit{\tera\electronvolt} \citep{aleksic2016major}, with a FoV of \SI{3.5}{\deg}. 
    Despite having a higher energy threshold than the LST-1, the MAGIC telescopes can achieve a better shower reconstruction (energy, direction and background suppression) thanks to the stereo trigger, which allows them to record only events that trigger both telescopes. 

    The inter-telescope distances, of around \SI{100}{\meter}, is comparable to the expected size of the Cherenkov light pool resulting from the extensive shower initiated by a VHE gamma ray. 
    Therefore, it is possible to combine LST-1 and MAGIC telescopes into a single stereoscopic system, improving the performance and the accuracy of event reconstruction~\citep{kohnle1996stereoscopic, abe2023performance}. 

    Observation campaigns searching for counterparts of GW events have been conducted following notably different procedures and selection criteria in the MAGIC Collaboration \citep{miceli2019following} and LST Collaboration \citep{carosi2021first}.
    Usually, the GW uncertainty region is much larger than the telescopes' FoV, and therefore a standard observation scheme is not feasible. 
    In the case of LST-1, the observation scheduling is obtained using the so-called \textit{tiling} strategy, relying on the \texttt{tilepy} code, described in \citet{seglar2024cross}. 
    \texttt{tilepy} enables the automatic scheduling of follow-up observations by determining the optimal sequence of coordinates to observe, in order to maximize the probability of detection and optimize observation conditions. 
    The scheduling is flexible and can be tailored to either cover the entire uncertainty region or focus only on the most probable areas, depending on the available observation time and configuration. 
    Instead of spending a fixed amount of time on a single set of coordinates, \texttt{tilepy} allows distributing that time fractionally across multiple potential positions. This tiling algorithm is being used by other IACTs (e.g. \citealt{abdalla2017tev}) since the beginning of the LIGO–Virgo observing run O2. 
    Similarly, tiling strategies have been consistently used in other wavelengths to address the poor localization of GW events (e.g., see \citealt{evans2016swift, ghosh2016tiling}). 
    Recently, the MAGIC and LST Collaborations have set up a joint GW program that enables automatic coordination of observation campaign scheduling, orchestrated by the Transient Handler of LST-1 \citep{seglar2025transients} using \texttt{tilepy}.
    
    GW240615 and GW241125 were extraordinary events because, due to the small uncertainty area and the GRB-like candidate identification respectively, the entire region of interest was contained within the telescopes' FoV. 
    Since these events occurred during daytime in La Palma, it allowed for offline coordination of the joint MAGIC+LST-1 observation campaign, marking the first time that observations were scheduled in a coordinated manner between IACTs for a GW event. 
    Given the low localization uncertainty, standard joint wobble observations were identified as the best strategy for both events. 
    In this configuration, the telescopes point alternatively at different directions at a fixed offset from the target \citep{Fomin1994}. 
    Typically, each pointing, or observing run, lasts for $\sim$\SI{20}{\minute}.    
    This mode allows to define \textit{off-source} regions within the FoV which have the same acceptance as the region around the source, hence facilitating background estimation and minimizing systematic effects related to reconstruction or camera response. 
    
    Standalone datasets are reduced with different analysis chains. 
    The standard pipeline for LST-1 data employs the software package \texttt{cta-lstchain v0.10.7}~\citep{Lopez-Coto2021performanceLST, abe2023performance, lstchain-Zenodo_2024}, while the MAGIC analysis is based on \texttt{MARS} \citep{moralejo2009mars, 2013ICRC...33.2937Z}. 
    To combine the two datasets and perform a stereo analysis, MAGIC data are converted into ctapipe-like data format using the \texttt{ctapipe$\_$io$\_$magic} package\footnote{\url{https://github.com/cta-observatory/ctapipe_io_magic}}. 
    The joint analysis is then performed using the \texttt{magic-cta-pipe} package\footnote{\url{https://github.com/cta-observatory/magic-cta-pipe}}. 
    In both cases, the package comes with additional modules for high-level analysis such as \texttt{pyirf} \citep{Noethe2022} for calculation of the instrument response functions (IRFs) and \texttt{gammapy} \citep{Donath2023gammapy, acero_2025_17814297_gammapy2.0} for flux modelling. 
    More details on the joint analysis pipeline can be found in \citet{abe2023performance}. 

\section{Observing conditions and data taking}\label{Sec:4}
    \subsection{Observations of the BBH system GW240615}

    We observed GW240615 as Target of Opportunity (ToO), with the MAGIC and LST-1 telescopes in stereo configuration. 
    Observations started \SI{14.88}{\hour} after the initial trigger from LVK. 
    LST-1 started observing the event at 2024-06-16 02:29:17 UTC and was followed by MAGIC with \SI{8}{\minute} delay. 
    The three telescopes observed consistent wobble positions (i.e. with a difference of pointings of less than \SI{0.1}{\deg}) for \SI{111.76}{\minute}, distributed in 6 observational runs. 
    The observations were simultaneously performed pointing to the first available set of coordinates (\textit{Bayestar} LVK reconstruction, 7.53, +45.81; \citealt{2024GCN.36669....1L}), although a better localization at \SI{13.49}{\arcminute} from the initial one was published a few days later (\textit{Bilby} LVK reconstruction, 7.69, +45.67; \citealt{2024GCN.36704....1L}).

    Observations were performed at a high zenith angle from \SI{65}{\deg} to \SI{42}{\deg}.  
    Additionally, data-taking was affected by variable night-sky-background (NSB) conditions, although weather conditions during the observations were optimal.  
    Both the level of NSB and the zenith angle have a significant impact on the energy threshold of the telescopes, thus a non-standard approach was necessary to characterize the observations of GW240615. 
    See Appendix~\ref{appendix:nsb-settings-e-threshold} for more information about the energy threshold and the NSB conditions of GW240615 and how we addressed them in our analysis.

    \subsection{Observations of the \textit{Swift}-BAT counterpart candidate of BBH system GW241125} 

    LST-1 observations of GW241125 started \SI{19.01}{\hour} after the initial trigger from LVK, at 2024-11-25 20:01:49 UTC, and continued for a total of \SI{4}{\hour}. 
    MAGIC observations started with a delay of \SI{29}{\minute}. 
    Observations were performed in wobble mode around the \textit{Swift}-BAT candidate coordinates (58.079, +69.689 deg; \citealt{2024GCN.38308....1D}). 
    In total, the three telescopes collected data in joint configuration for \SI{201.56}{\minute}. 
    The zenith angle of the observations went from \SI{55}{\deg} to \SI{40}{\deg}. 
    
\begin{figure}[]
        \includegraphics[width=1\columnwidth]{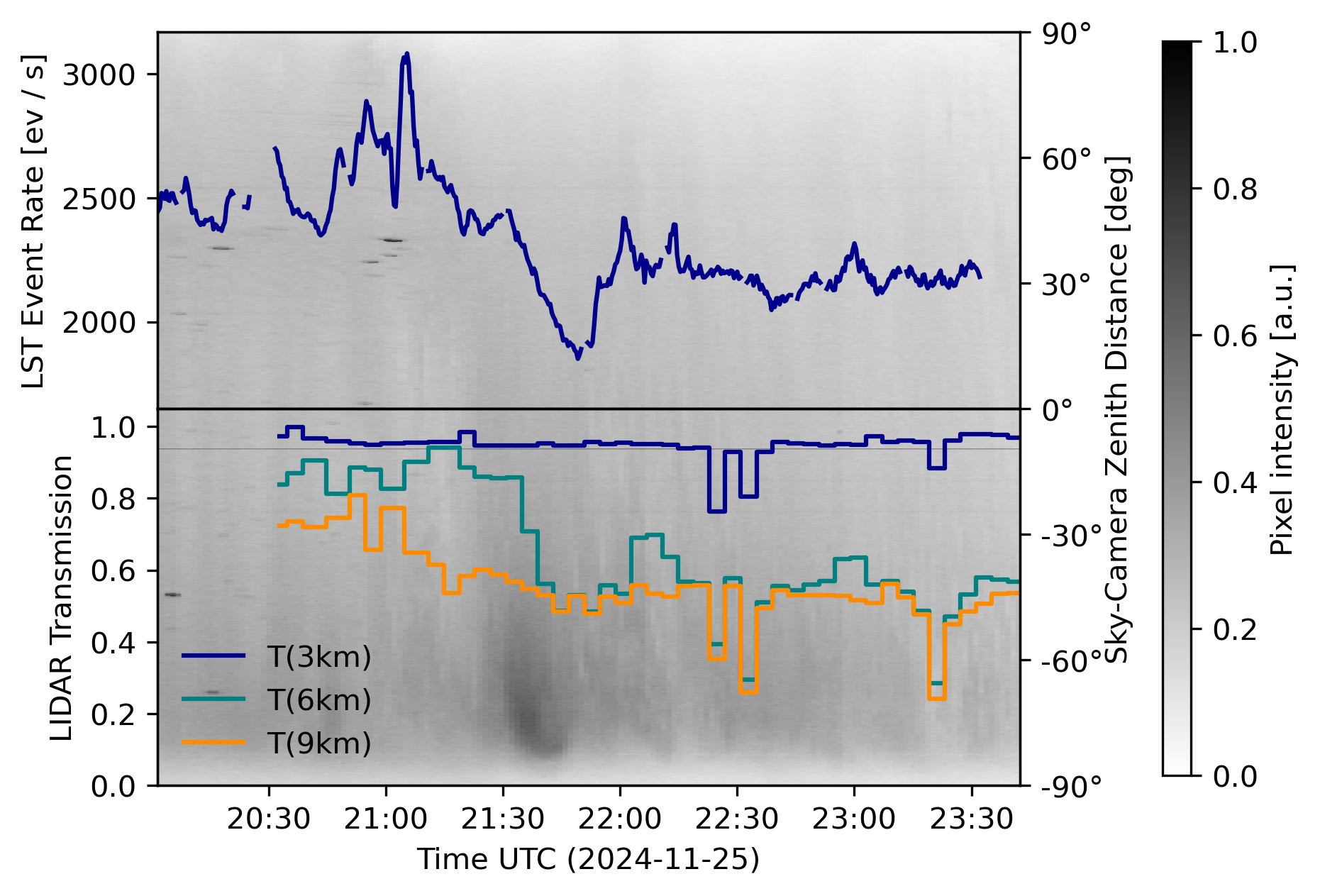}
        \caption{The top plot shows the LST-1 event rates over the night. The bottom plot shows the atmospheric transmission from different altitudes at ground level as determined by the MAGIC LIDAR for the observation of GW241125. The keogram taken at the same time at GTC ($\sim$ \SI{500}{\meter} away from MAGIC) is displayed in the background, with the zenith distance angle of the SkyCamera position. Clouds were present throughout the night, with particularly thick layers appearing after 21:30 UTC. These led to a sharp drop in atmospheric transmission, degrading the data quality to the point where meaningful analysis was no longer possible.}
        \label{fig:keogram}
    \end{figure}

    The observation conditions were suboptimal. 
    A Light Detection and Ranging (LIDAR) system has been deployed by the MAGIC collaboration for the monitoring of atmospheric conditions during regular observations\citep{Schmuckermaier2023CorrectingSystem}. 
    During the observations of GW241125, LIDAR measured vertical atmospheric transmission from \SI{9}{\kilo\meter} above ground level decreasing from 0.8 to 0.4 due to a persistent layer of clouds. 
    The clouds left only very narrow and sporadic windows of clear sky throughout the night.
    Due to these poor atmospheric conditions, the data could not be analysed using standard methods and most of them were excluded from the final analysis.
    To better assess atmospheric quality, we examined the LIDAR and the \textit{keogram} data from that night. 
    The keogram is a plot showing the temporal evolution of brightness along a fixed image slice (usually north–south) from All-Sky camera images (Fig.~\ref{fig:keogram}). 
    In this case, the keogram was provided by the GTC All-Sky camera\footnote{\url{https://atmosportal.gtc.iac.es/}}.
    Both monitoring tools show consistent behaviour: the atmospheric transmission at \SI{6}{\kilo\meter} measured by LIDAR drops simultaneously with the appearance of a thick cloud in the keogram, indicating a sudden deterioration of atmospheric conditions.
    We select the windows with higher transmission resulting in a total of \SI{19.72}{\minute} of joint data and \SI{37.68}{\minute} of LST-1 data, and we set a conservative energy cut of \SI{0.5}{\tera\electronvolt}. 
    A detailed description of the cut selection procedure can be found in Appendix~\ref{appendix:nsb-settings-e-threshold}.
    
\section{Methodology}\label{Sec:5}

    \subsection{Low-level analysis}
    
    We analysed the data taken with LST-1 in standalone mode using \texttt{lstchain} \citep{Abe2023} as the main reconstruction framework, and \texttt{magic-cta-pipe} \citep{abe2023performance} for the stereo reconstruction. 
    The procedure starts from the measured charge values, which are corrected, integrated, and cleaned to produce calibrated images. 
    These images are then parametrized, producing Hillas parameters \citep{1985ICRC....3..445H_hillas_parameters}. 
    A random forest algorithm is applied to estimate the event energy, arrival direction, and a classification parameter (\textit{gammaness}) that separates gamma-like events from background-like events.
    
    \subsection{Sources with uncertain localization}\label{Subsec:sources-uncertain-loc}
    
    We describe here the methodology used to analyse sources with uncertain localization, specifically in cases where the positional uncertainty exceeds the point spread function (PSF) of the telescope. 
    This approach is applied to the BBH event GW240615, which has not been associated with any known galaxy.
    
    The estimation of the background is critical for IACTs: even after the selection of gamma-like events, the rate of residual background from showers initiated by charged cosmic rays is typically comparable or larger than the rate of gamma rays from the majority of sources. 
    We used the background modelling tool \texttt{pybkgmodel}\footnote{\url{https://github.com/cta-observatory/pybkgmodel}} \citep{CTA-LSTproject:2023ovx_pybkgmodel}. 
    A cross-check of the results was performed using a different tool, \texttt{BAccMod} \footnote{\url{https://github.com/mdebony/BAccMod}}. 
    Background is estimated using a "3D" model (not assuming radial symmetry), and computed run-wise. 
    In addition, we used the \textit{ring background method} \citep{2007A&A...466.1219B_bkg_modelling_gamma} to normalize the model over the dataset. 

    After background modelling, we defined the respective ON and OFF regions to compute Li\&Ma significance \citep{1983ApJ...272..317L_li_and_ma_significance} and the flux estimators \citep{1969NucIM..70..200H}. 
    In the flux calculations, we set a minimum flux of 0 allowed to likelihood estimation. 
    We did not consider negative statistical fluctuations for the flux computation (see details in Appendix~\ref{appendix:gammapyModifications}). 
    
    Finally, we used the \texttt{ExcessMapEstimator} tool provided by the \texttt{gammapy} framework to generate the final results (significance and U.L.s) in the form of a sky-map. 
    Since the output represents the integrated flux within a specific correlation radius, we applied a correction factor to account for the fraction of the signal of a point-like source contained within that region (see Appendix~\ref{appendix:gammapyModifications}). 
    For GW240615, we chose correlation radii consistent with the PSF of the instrument: specifically, we used \SI{0.25}{\degree} for the low-energy analysis and \SI{0.1}{\degree} for the higher-energy analysis, both described in Sec.~\ref{Sec:6}. 
    
    \subsubsection{Global statistical measures using GW localization}\label{Subsec:general-parameters-est}
    
    We can combine the information from the LVK observation (the 2-D sky-map of the GW event localization) with the gamma-ray data to obtain a single "global" value for the significance, and another one for the flux upper limit (instead of the sky-maps of such quantities). 
    Several approaches have been proposed to address this problem: some rely on a Bayesian framework \citep{Abbasi_2023_IceCube2023_bayesian_and_binned_lik, PhysRevD.100.083017_multimessenger_bayesian, Veske_2020_neutrino_bayesian}, others adopt a frequentist perspective \citep{PhysRevD.85.103004_multimessenger_frequentist, Hussain2019}, while a third class of methods takes a hybrid approach combining elements of both \citep{Vianello_2017_fermi_gw_significance}.

   \begin{figure*}[t]
        \includegraphics[width=1\linewidth]{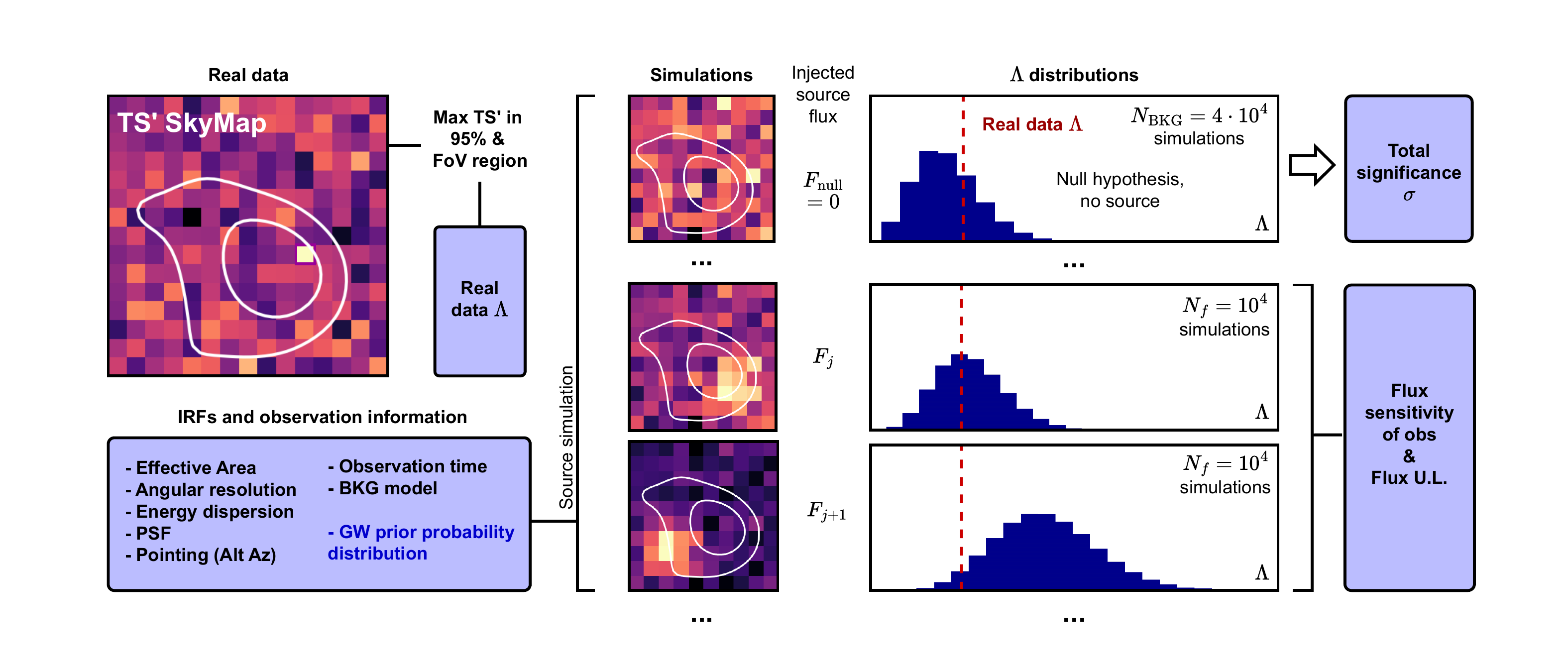}
        \caption{Schematic representation of the frequentist parameter estimation workflow. 
        From left to right: (1) Real Data: Gamma-ray data (skymap) is combined with the GW spatial prior (white contours) to generate a modified test statistic map, $TS'$. 
        The maximum value within the overlap of the 95$\%$ GW localization and the instrument FoV is defined as the observed statistic, $\Lambda_{\rm data}$. 
        (2) Simulations: Toy MC realizations are generated using the relevant IRFs and observation metadata. 
        These include a null-hypothesis set ($F_{\rm null}=0$) and multiple sets with injected source fluxes ($F_{\rm j}$) sampled from the GW prior.
        (3) $\Lambda$ distributions: The resulting probability density functions of $\Lambda$ for each simulation set are compared against $\Lambda_{\rm data}$ (red dashed line). 
        (4) Results: These comparisons yield the total significance ($\sigma$), global flux U.L., and the flux sensitivity of the specific observation.
        See text for more details.}
        \label{fig:diagram-method-frequetist}
    \end{figure*}
    
   For this project, we adopted a frequentist approach following \citealp{Hussain2019} (see the full scheme in Fig.\ref{fig:diagram-method-frequetist}). 
   We modify the standard test statistics (TS) used for significance computation \citep{1983ApJ...272..317L_li_and_ma_significance}, which for each pixel $i$ is defined as the Likelihood Ratio Test between the two Poisson likelihood hypothesis:
    \begin{equation}
        \text{TS}_i=2\log\left(\frac{\mathcal{L}_i^{\text{source+BKG}}}{\mathcal{L}_i^{\text{BKG}}}\right).
    \end{equation}
   
    Following \citet{Hussain2019}, we define the modified test statistic $TS'$ by weighting the likelihood with the GW prior spatial information in each pixel (i.e. the 2-D probability map of GW localization). 
    In this way, we are also incorporating the information from the LVK likelihood by integrating it over each pixel. 
    Since the GW and gamma-ray data are independent, and the standard likelihood is already integrated in the first part, the combined likelihood can be expressed directly as the product of the individual likelihoods. 
    We use WCS geometry \citep{Donath2023gammapy, 2003ASPC..295..403G_coordinate_system_fits}, with square (in RA-Dec) bins with slightly different solid angle between them. 
    The LVK sky-map of probabilities is provided in HEALPix coordinates \citep{Górski_2005_healpix}. 
    The integrated probability per WCS pixel, $P_i^{GW}$, is obtained by summing the HEALPix pixels contained within each WCS pixel and normalizing by their areas. 
    With a resolution of about 12 HEALPix pixels per WCS pixel, the transformation maintains fidelity to the original map.

    \begin{equation}
        \text{TS}'_i = 2\log\left( \frac{\mathcal{L}_i^{\text{source+BKG}}\cdot P_i^{GW}}{\mathcal{L}_i^{\text{BKG}}} \right)=\text{TS}_i+2\log\left(P_i^{GW}\right).
        \label{eq:ts'}
    \end{equation}

    The search of the largest $TS'$ is then restricted to the overlap between:
    
    \begin{enumerate}
        \item The \SI{95}{\percent} containment region of the GW localization, based on the LVK prior. 
        Although some works (e.g., \citealt{Vianello_2017_fermi_gw_significance}) use \SI{90}{\percent}, we adopt \SI{95}{\percent} to maximize the observed data used in the analysis.
        \item The region covered by the observations. 
        Different criteria may be adopted, such as a constant search radius or an exposure threshold. 
        In the present case, since the entire \SI{95}{\percent} region is above the 30th percentile of the nonzero exposure distribution, we consider that is well covered (see more details in Appendix~\ref{appendix:nsb-settings-e-threshold}, Fig.\ref{fig:maps-exposure}).
    \end{enumerate}

    Within this overlap region, we identify the maximum value of $\text{TS}'$, which we denote as the derived test statistic, $\Lambda$. 
    To understand the distribution of $\Lambda$ under the null hypothesis, we use toy MC simulations produced with \texttt{gammapy} \citep{Donath2023gammapy, acero_2025_17814297_gammapy2.0}. 
    These simulations use the same IRFs and background (BKG) models as the analysis, along with the pointing information and time and date of observations. 
    Comparing the observed $\Lambda$ to the null distribution for $N_{\text{BKG}}$ simulations (in our case, we used $N_{\text{BKG}}=40000$), the p-value is computed as the fraction of events with larger $\Lambda$ than the observed $\Lambda_{\text{data}}$.
    The significance (in units of $\sigma$) is, instead, estimated using the inverse cumulative distribution function of the standard normal distribution: $\Phi^{-1}(1 - \text{p-value})$.
    
    In addition, multiple simulations are produced for a range of flux values, with $N_f=10000$ realizations for each. 
    Each realization consists of a toy MC injecting a source at a random sky position drawn from the GW probability distribution. 
    In our analysis, the source was modelled by a power-law with spectral index $-2$. 
    For each value of amplitude, we computed $N_f$ realizations. 
    We choose amplitude values equally spaced in logarithmic scale, with 40 bins per decade, from \num{e-14} to $\num{e-9}$ erg \SI{}{\per\square\centi\metre\per\second}. 
    We ignore the cases where the simulated source falls outside the \SI{95}{\percent} containment region.
    Consequently, if only a fraction $F$ of the GW localization region is observed, our results carry an inherent uncertainty of $1-F$. 
    The reported upper limits are therefore conditional on the assumption that the GW source lies within the observed region.

    We aim to determine the sensitivity of the observations, i.e. the minimum flux that would have been detected by the method for a given set of observations. 
    Following a similar approach to other works \citep{Vianello_2017_fermi_gw_significance, Hussain2019}, we define the U.L. as the flux for which \SI{95}{\percent} of trials yield a $\Lambda$ value larger than the one of the real data, i.e. $\Lambda_f > \Lambda_{\text{data}}$.
    If the observed $\Lambda_{\text{data}}$ is smaller than $\Lambda_{\text{bkg-median}}$, the median of the null hypothesis distribution, or equivalently, if the total significance is negative, this indicates a downward fluctuation of the background: in such cases, we take a conservative approach. 
    We define the sensitivity of our observations as the flux for which \SI{95}{\percent} of trials yield a $\Lambda$ value larger than $\Lambda_{\text{bkg-median}}$, i.e. $\Lambda_f > \Lambda_{\text{bkg-median}}$. 
    We use the sensitivity of the observations as a U.L. in those cases. 
    The global U.L.s presented have therefore a \SI{95}{\percent} confidence level (CL), in addition to the implicit assumption of the source being located in the \SI{95}{\percent} GW probability region, resulting on a global CL of \SI{90}{\percent}. 
    
    \subsection{Analysis of localized sources in suboptimal weather conditions (GW241125)}

   \begin{figure*}[t]
        \includegraphics[width=1\linewidth]{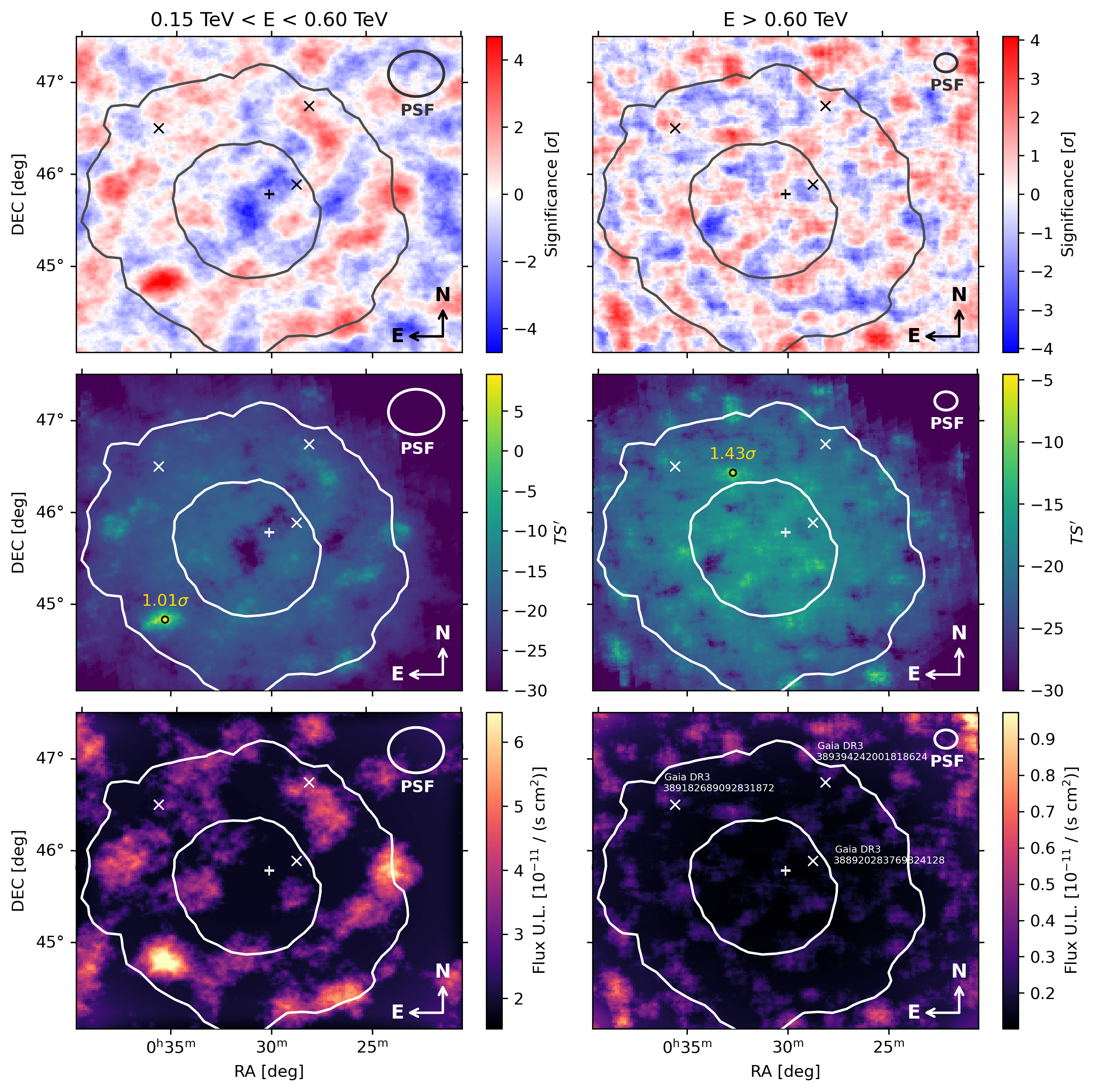}
        \caption{The top row shows the significance sky-maps for GW240615, analysed in both energy ranges. The middle row is showing the $TS'$ maps (eq. \ref{eq:ts'}) and the largest significance point of the interest region (with black circles), superposing the resulting global significance on them.
        The flux U.L. sky-maps for both energy ranges with \SI{95}{\percent} CL is shown in the bottom row. 
        The contours show \SI{50}{\percent} and \SI{95}{\percent} GW containment region, while the cross corresponds to the Bayestar position, i.e. the direction followed by our observations. 
        The PSF (\SI{68}{\percent} containment) averaged over the energy range is indicated in the top-right of each panel.
        The "x" signs represent the position of the 3D coincident AGNs in the catalogue crossmatch (See Sect.\ref{Sec:7}).
        }
        \label{fig:maps-flux}
    \end{figure*}
    
    As discussed in Sect.~\ref{Sec:5} the observations of GW241125 were affected by suboptimal weather conditions, including a thick layer of clouds, as indicated by LIDAR measurements and by telescope rates, that are $\sim$\SI{60}{\percent} lower than optimal values (Fig. \ref{fig:keogram}). 
    Most of the data were discarded for quality reasons, but also the remaining dataset is far from optimal, with event rates $\sim$ \SI{40}{\percent} lower than in normal, clear and dark-sky conditions. 
    To maximize the scientific return from these observations, we performed a tailored analysis specifically optimized for the atmospheric conditions during the observations.

    The atmosphere, while essential to the detection process, is also an important source of systematic uncertainties. 
    The extinction of Cherenkov radiation due to scattering and absorption by aerosols and cloud layers can significantly reduce the number of photons reaching the telescopes, thereby decreasing the Cherenkov light yield and the effective collection area \citep{Bernlohr2000ImpactTechnique, Sobczynska2014InfluenceEnergies, Pecimotika_2023}. 
    Consequently, if these effects are not taken into account, a systematic underestimation of both the reconstructed primary gamma-ray energy and flux may occur \citep{Schmuckermaier2023CorrectingSystem, Zywucka2024ANovel}.

    We generated adapted MC simulations following the approach presented in \citet{Pecimotika_2023}. 
    The central idea of this approach is to calibrate the telescope response according to the prevailing atmospheric conditions. 
    To this end, MC simulations are produced using dedicated atmospheric transmission profiles, specifically incorporating cloud properties derived from measurements with atmospheric monitoring instruments such as the LIDAR system. 
    This allows to model the impact of clouds on the Cherenkov signal detected in the telescope camera. 
    As shown by \citet{Pecimotika_2023} and \citet{Zywucka2024ANovel}, using adapted MC simulations to generate random forest models reduces the energy bias to the levels seen in clean atmosphere observations, without impairing energy resolution.

    MC simulations of extensive air showers induced by primary gamma rays and protons were generated using CORSIKA \citep{Heck1998CORSIKA:Showers}, while the subsequent propagation and extinction of Cherenkov photons through the atmosphere, along with the detector response, were simulated using the \texttt{sim\_telarray} code \citep{Bernlohr2008SimulationSim_telarray}. 
    The table of optical depths needed for calculations with \texttt{sim\_telarray} was generated using the MODerate spectral resolution atmospheric TRANsmittance (MODTRAN) band model algorithm \citep{Berk1987MODTRAN:LOWTRAN}.  
    The baseline (clear-sky) atmospheric transmission profile was adapted from the default atmospheric configuration used in the LSTProd2\footnote{\url{https://github.com/cta-observatory/lst-sim-config}} settings for \texttt{sim\_telarray}, namely using the tropical atmosphere with navy maritime extinction. 
    The cloud layer was integrated into the baseline profile by specifying cloud parameters in the MODTRAN simulations, based on measurements from MAGIC's elastic micro-Joule LIDAR \citep{Fruck2022CharacterizingLIDAR}. 
    The cloud layer for GW241125 observations was characterized by a vertical transmittance of 0.736, a base height of \SI{6.86}{\kilo\meter} above sea level, and a total thickness of \SI{3.68}{\kilo\meter}, assuming wavelength-independent and uniform extinction throughout the cloud \citep{Fu1996AnModels, Serrano2015WavelengthDepth}. 
    General settings for simulations with CORSIKA and \texttt{sim\_telarray} for both joint and mono observations are described in \citet{abe2023performance, Abe2023}, respectively.
    
\section{Results}\label{Sec:6}
    
    \subsection{GW240615}    
    
    We divided the data analysis of GW240615 in two energy ranges to account for the varying energy threshold and fully exploit the capabilities of the LST-only and MAGIC+LST-1 configurations (see Appendix~\ref{appendix:nsb-settings-e-threshold}).
    In particular, LST-only data are analysed in the low-energy range \SI{0.15}{\tera\electronvolt} $< E_\gamma <$ \SI{0.6}{\tera\electronvolt}, while MAGIC+LST-1 in the high-energy range, $E_\gamma >$ \SI{0.6}{\tera\electronvolt}. 
    
    For the gamma–hadron separation efficiency --- that is, the energy-dependent cuts applied to data and MC simulations to retain a given fraction of gamma-ray events after reconstruction --- we adopted values of \SI{50}{\percent} for the low-energy analysis and \SI{70}{\percent} for the high-energy analysis.
    The higher efficiency needed in the high-energy range reflects the improved background suppression provided by stereo observations.
    
    We computed statistical significance sky-maps (see Fig.~\ref{fig:maps-flux} and Appendix~\ref{appendix:SignificanceMaps}); no significant regions are found above $5\sigma$.
    Using the method described in Sect.~\ref{Subsec:general-parameters-est}, we obtained global significance estimates of \num{1.01}$\sigma$ for low-energy analysis and \num{1.43}$\sigma$ for high-energy analysis (see also Fig.\ref{fig:maps-flux}). 

\begin{figure}[H]
        \includegraphics[width=1\columnwidth]{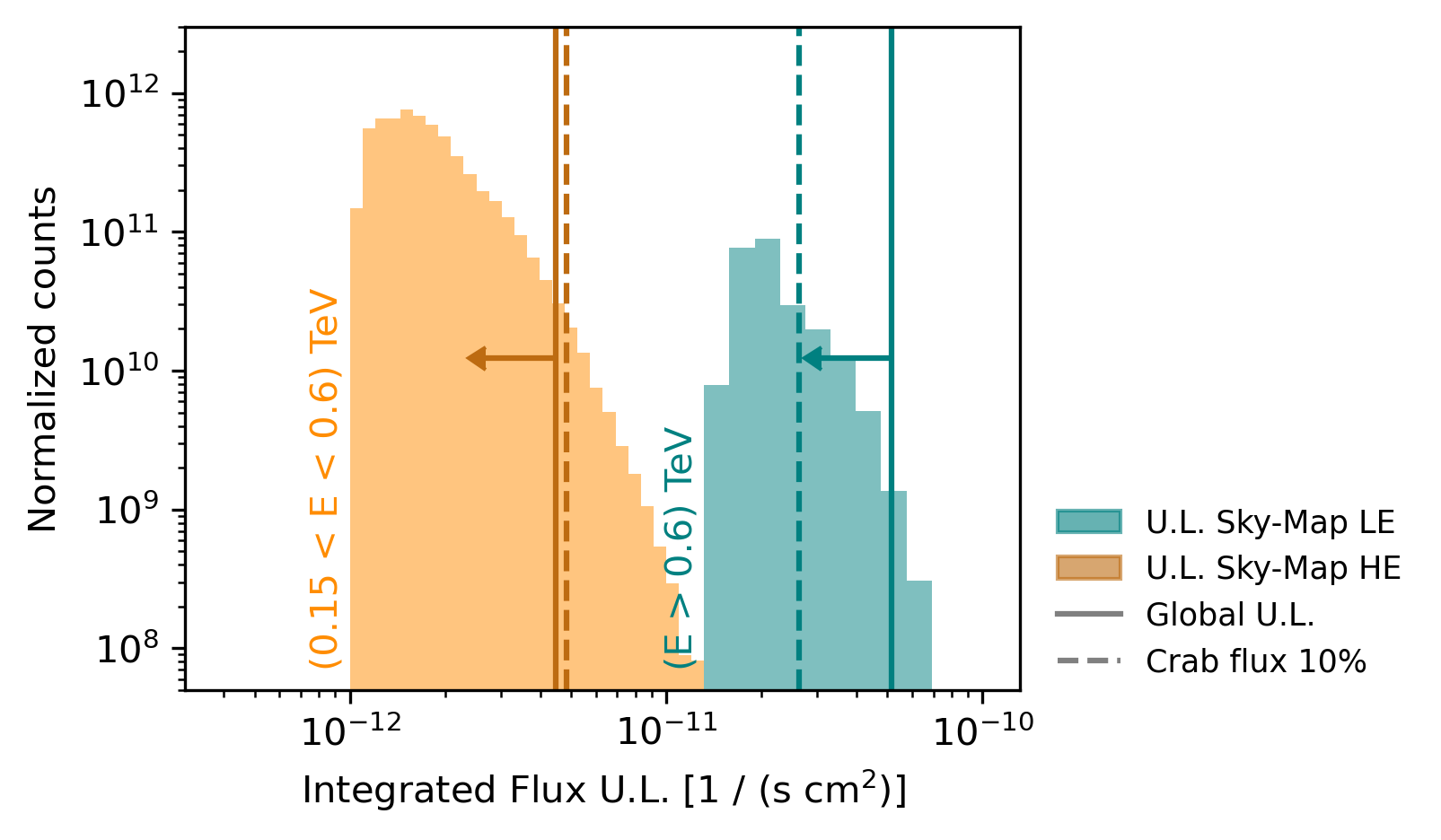}
        \caption{Flux upper limits of GW240615 in the $0.15 < E < 0.6$ TeV (teal) and $ E > 0.6$ TeV (orange) energy ranges are shown. 
        The histograms represent the sky-map U.L.s distribution for all the pixels. 
        Vertical solid lines represent the global U.L., derived from the uncertain source localization method described in Sect.~\ref{Subsec:sources-uncertain-loc}. 
        Vertical dashed lines represent the \SI{10}{\percent} of Crab in the corresponding energy range (from \citealt{2015JHEAp...5...30A_MAGIC_CrabSpectrum}).}
        \label{fig:flux-2d-projection}
    \end{figure}

    We also computed the flux sky-maps (Fig.\ref{fig:maps-flux}). 
    All the U.L.s in the sky-map are produced with \SI{95}{\percent} CL. 
    No significant structures or excesses are visible, and the maps are consistent with statistical fluctuations, in line with the previous significance results. 
    As expected, the upper limits are higher in the regions where the instrument is less sensitive, such as at the edges of the field of view. 
    The corresponding projected distributions for both sky-maps are presented in Fig.~\ref{fig:flux-2d-projection}, which also includes the derived flux U.L.s, together with the integrated Crab flux in the energy ranges (from \citealt{2015JHEAp...5...30A_MAGIC_CrabSpectrum}). 
    We computed the global flux U.L.s with the method detailed in Sect.~\ref{Subsec:general-parameters-est}. 
    Results are summarized in Table~\ref{tab:ULs}. 

\begin{table}[htbp]
\centering
\caption{Flux upper limits of GW240615.}
\label{tab:ULs}
\begin{tabular}{cccc}
\hline
\hline
Energy Range & Significance & U.L. & U.L. \\
(TeV) & ($\sigma$) & ($\text{cm}^{-2}\,\text{s}^{-1}$) & (\% C.U.) \\
\hline
$0.15 < E < 0.6$  & $1.01$ & $5.17\times10^{-11}$ & 19.7 \\
$E > 0.6$      & $1.43$ & $4.47\times10^{-12}$ & 9.2\\
\hline
\end{tabular}
\raggedright
\tablecomments{Upper limits are computed for the LST-1 ($0.15 < E < 0.6$ TeV) and for the MAGIC+LST-1 ($E > 0.6$ TeV) energy ranges. We used a confidence level of 95\%. Values are reported in $\text{cm}^{-2}\,\text{s}^{-1}$ and percentages of Crab Units (\% C.U.).}
\end{table}

    \subsection{GW241125}

   For the analysis of GW241125, where a candidate counterpart was identified \citep{2024GCN.38305....1L, 2024GCN.38313....1L, 2024GCN.38308....1D}, no sky-map search was performed. 
   The LVK uncertainty without the candidate counterpart information was too large enough (\SI{2196}{\square\deg}) to be observed with IACTs, hence we only observed the small region where the candidate counterpart was located. 
   We computed the significance, the Spectral Energy Distribution (SED) and the Light Curve (LC) treating the GW as a point-like source. 
   The entire analysis was performed using the standard \texttt{gammapy} tools \citep{Donath2023gammapy, acero_2025_17814297_gammapy2.0}. 
   The CL of the U.L.s provided both in the SED and the LC are \SI{95}{\percent}.
    
\begin{figure}[H]
        \includegraphics[width=1\columnwidth]{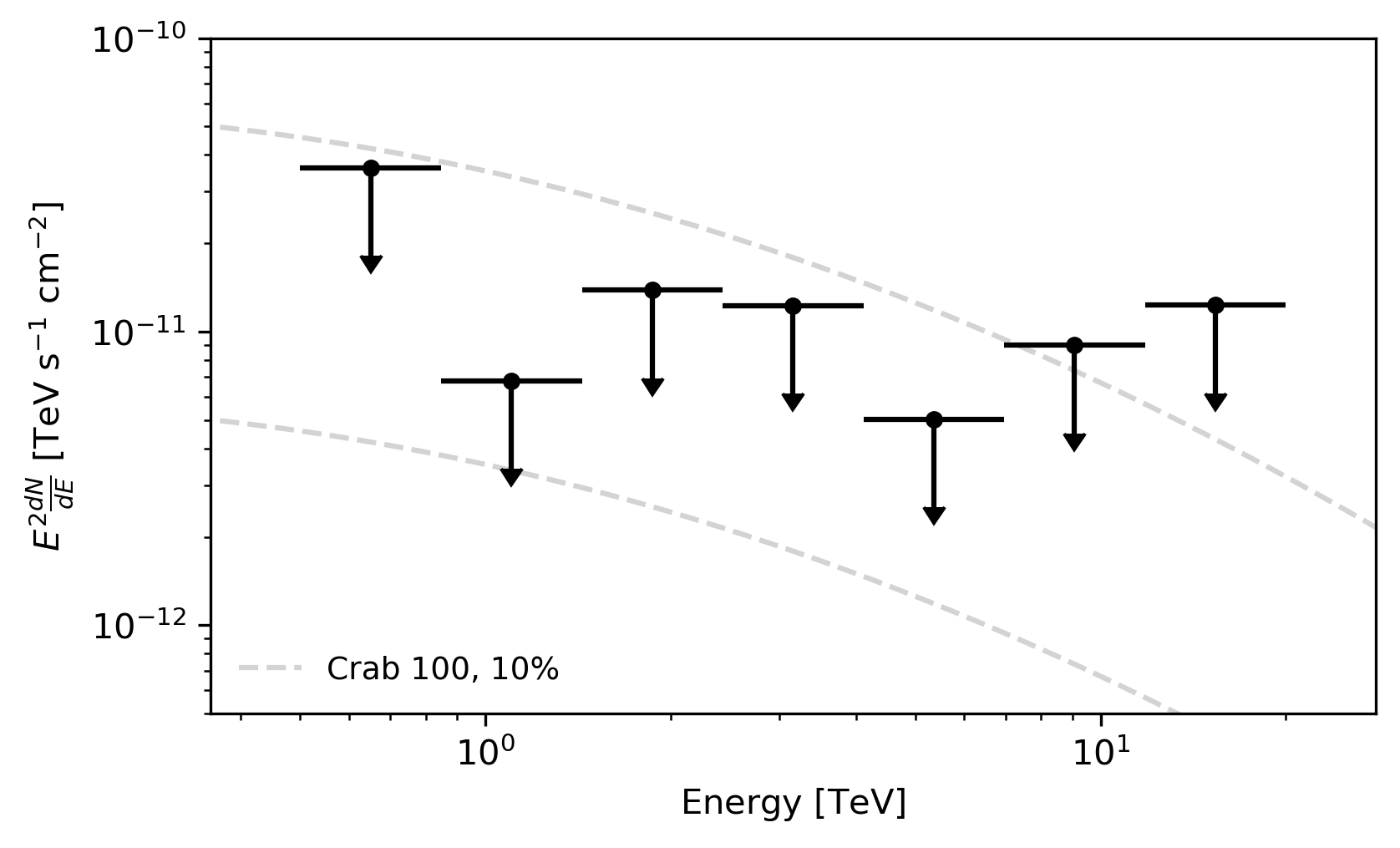}
        \caption{SED calculated for the candidate counterpart position of GW241125. 
        The source was not statistically detected in any of the energy bins. 
        The \SI{100}{\percent} and \SI{10}{\percent} MAGIC Crab SED are also plotted for comparison \citep{2015JHEAp...5...30A_MAGIC_CrabSpectrum}}
        \label{fig:SED-lst-magic}
    \end{figure}

\begin{figure}[H]
        \includegraphics[width=1\columnwidth]{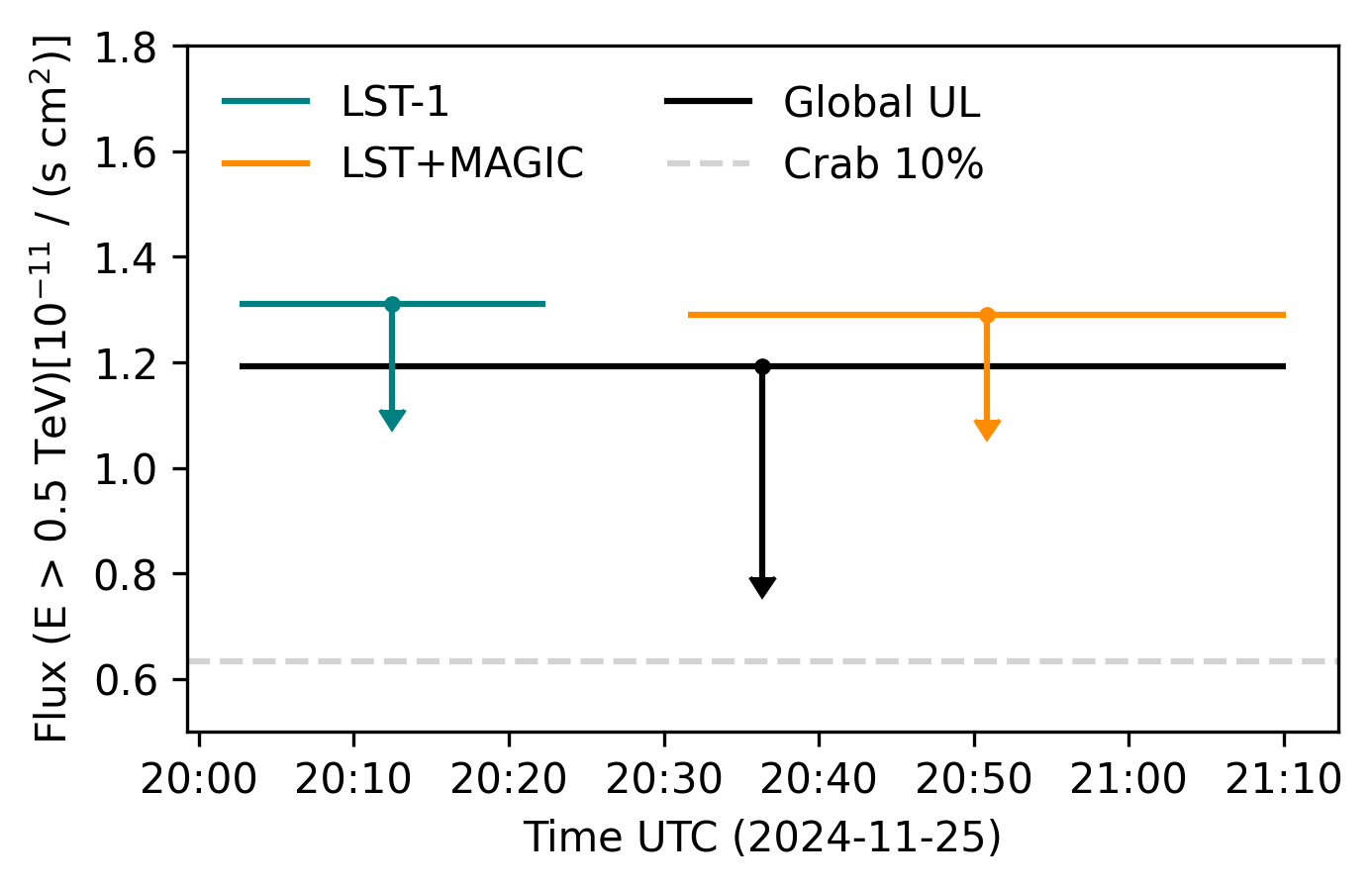}
        \caption{The LC of the observations of GW241125 for energies larger than \SI{0.5}{\tera\electronvolt}. 
        It has been separated in 3 runs of about \SI{20}{\minute}, where we distinguish between the run that was taken with LST-1 in mono mode and the ones taken in stereo mode with MAGIC+LST, that are shown as a unique limit in the plot. 
        We also show the global U.L. using stacked data.}
        \label{fig:LC-lst-magic}
    \end{figure}

    We computed the distribution of separations of reconstructed events to the source position for ON and OFF regions (see Appendix~\ref{appendix:SignificanceMaps} and Fig.~\ref{fig:theta-plot} for more details) obtaining statistical significance of \num{2.2}$\sigma$ for LST data and \num{1.47}$\sigma$ for LST+MAGIC data.
    The SED and the LC are computed for reconstructed energies $E>$ \SI{0.5}{\tera\electronvolt} and are shown in Fig.~\ref{fig:SED-lst-magic} and Fig.~\ref{fig:LC-lst-magic}, respectively. 
    The light curve was computed using a power-law spectral model with a fixed index of $\alpha = -2$. 
    Since the first observation run was performed with LST-1 alone, the resulting limits are expected to be weaker, reflecting the reduced sensitivity of mono reconstruction compared to the stereo reconstruction. 
    The differences seen between LC points, correspond to statistical fluctuations: lower significances yield tighter U.L.s and vice-versa. 
    Overall, the derived limits are consistent with each other.  
    From approximately \SI{1}{\hour} of observations, combining the individual likelihoods, we obtain a global flux upper limit of \SI{1.2e-11 }{\per\second\per\square\centi\meter}.
    
\section{Discussion}\label{Sec:7}

    The two sources presented in this paper offered an unprecedented opportunity to observe a compact binary coalescence with gamma-ray ground-based telescopes. 
    Although no gamma-ray signal was detected, the derived upper limits allow discussing current theoretical frameworks that predict a detectable VHE flux. 
    In this regard, we exclusively focus on models of BBH mergers embedded within AGN disks, as these environments offer the necessary conditions for accelerating particles to VHE energies, unlike the case of isolated systems.
    Specifically, we discuss two models, both of them requiring accretion and subsequent jet launching, and predicting strong EM emission temporally coincident with the GW signal.
    We also consider the implications for future observational facilities and the observability by current telescopes. 
    Additionally, we explore AGN association.
    It is important to note that a detailed theoretical interpretation of these events is beyond the scope of this manuscript. 
    Therefore, our discussion relies on strong assumptions and physical conditions proposed in the existing literature.

\subsection{Super-Eddington accretion}\label{sec:accretion}
     
    Accretion of gas onto a BH in a binary system is a well established mechanism that allows the production of an EM signature, detectable at high energies \citep{shakura1973black, abramowicz2013foundations, belloni2016transient, bambi2024handbook}. 
    If the rate of accretion is larger than a critical value $\dot M_{\text{Edd}}$, corresponding to a limit luminosity $L_{\text{Edd}}$, the radiation force will exceed the gravitational force, driving fast and high energetic outflows \citep{jiang2014global, mckinney2014three, murase2016ultrafast}. 
    One of the most compelling evidence of super-Eddington accretion are the ultra-luminous X-ray binaries, which can exceed by 10 -- 1000 times the Eddington limit depending on the mass of the compact object (e.g.~ \citealt{begelman2002super, farrell2009intermediate, pinto2023ultra}). 
    Given that the accretion mainly depends on the presence and properties of a gaseous disk around the BH, a similar mechanism is proposed for BBH systems embedded in the accretion disk of an AGN (e.g.~\citealt{stone2017assisted, bartos2017rapid}). 
    In this environment, the merger is accelerated by loss of angular momentum of the binary due to viscous friction with the gaseous disk, resulting in mergers in less than \SI{1}{\mega\year} \citep{bartos2017rapid}. 
    The binary may accrete significant amount of gas from the AGN disk, which can exceed the Eddington limit producing outbursts.     
    In proximity of the merger, the escaping flux can be enhanced by several orders of magnitude compared to similar mass BHs thanks to shocks produced by the orbital motion of the binary \citep{farris2010binary}. 
    In particular, \citet{bartos2017rapid} discussed a geometrically thin, optically thick, radiatively efficient and steady-state accretion disk, as expected in AGNs \citep{shakura1973black}. 
    If the BBH is buried within the disk, emission can be reprocessed, effectively losing energy to the disk. 
    This would result in an optical/IR signature spread in time. 
    If, instead, during the spiralling phase, the binary has created some open gaps within the AGN disk (e.g.  \citealt{duffell2015simple, bartos2017rapid}), the high-energy emission can pass through and be detected on Earth. 
    The timescale of this emission varies from hours to days depending on the mass of the remnant, the accreting mass along the orbit of the BH and the mass accretion rate (e.g. \citealt{murase2016ultrafast}). 
    \citet{bartos2017rapid} defined the detectability of high-energy emission from the accreting binary by modelling the observed flux as a function of the super-Eddington efficiency.
    They define: 
    \begin{equation}
    \Phi_\gamma = 10^{-14} \eta_{\rm bol} \epsilon_\gamma \left(\frac{M_{\text{tot}}}{100 M_\odot}\right)\left(\frac{D}{100\,\text{Mpc}}\right)^{-2} \, \mbox{erg s$^{-1}$ cm$^{-2}$}
    \label{eq:1}
    \end{equation}
    where $\Phi_\gamma$ is the observed flux in gamma-rays, $\eta_{\rm bol} = L_{\rm bol}/L_{\text{Edd}}$ is the super-Eddington efficiency, $\epsilon_\gamma = L_{\gamma}/L_{\text{bol}}$ is the fraction of the bolometric luminosity $L_{\text{bol}}$ emitted in gamma-rays.
    Note that \citet{bartos2017rapid} discussed the observability of such emission solely within the energy ranges of the \textit{Chandra} and \textit{Fermi} space telescopes. 

\begin{figure}
    \includegraphics[width=1\columnwidth]{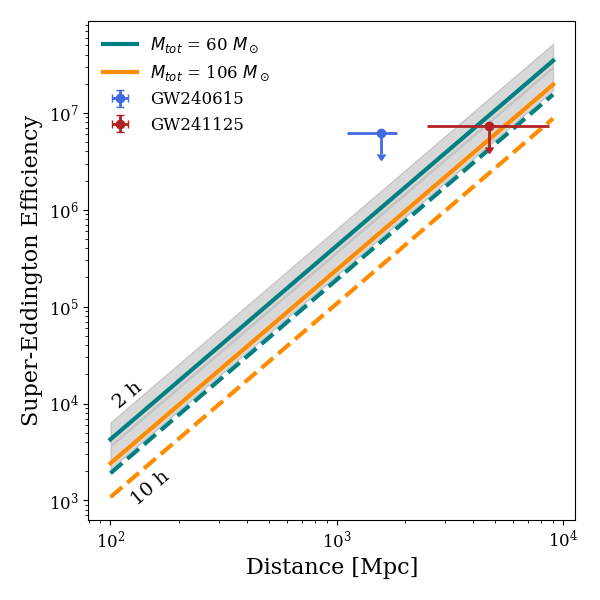}
    \caption{Super-Eddington efficiency $\eta_{\rm bol}$ evolution with distance. 
    Diagonal lines represent the efficiency lower limits considering the sensitivity of MAGIC+LST-1 at 500 GeV with 2 hours (thick solid lines) and 10 hours (dashed lines) of integration at zenith angle $<$ \SI{30}{\deg} \protect\citep{abe2023performance}.
    The gray bands cover the sensitivity area between $0.3 -$\SI{8}{\tera\electronvolt}. 
    Efficiency is estimated with two possible $M_{\rm tot}$, 60\,M$_\odot$ (orange) and 106\,M$_\odot$ (green), and assuming a partition factor of $\epsilon_\gamma$ = 0.1.
    For comparison, we plot the GW240615 ($M_{\text{tot}} \sim 60.4 M_\odot$ and $D_{\rm L} = 1560^{+290}_{-450}\,\rm Mpc$) and GW241125 ($M_{\text{tot}} \sim 106 M_\odot$ and $D_{\rm L} = 4700^{+3900}_{-2200}\,\rm Mpc$) upper-limits. 
    }
    \label{fig:eff}
    \end{figure}

    Using Eq.~\ref{eq:1} with the masses, the distances and the flux upper-limits of GW240615 and GW241125, and assuming $\epsilon_\gamma = 0.1$, we constrain the super-Eddington efficiency of these sources.
    We obtained $\eta_{\rm bol} \le$ \num{6.3e6} and $\eta_{\rm bol}\le$ \SI{7.4e6}, respectively. 
    Additionally, we use the sensitivity curves of the MAGIC+LST-1 configuration to put a lower-limit on the super-Eddington efficiency necessary to detect a gamma-ray signature with our telescopes. 
    With 2 and 10 hours of exposure at \SI{500}{\giga\electronvolt}, the MAGIC+LST-1 are able to reach a flux of $\sim$ \SI{2.6e-12}{\erg\per\square\centi\meter\per\second} and $\sim$ \SI{1.2e-12}{\erg\per\square\centi\meter\per\second}, respectively \citet{abe2023performance}.  
    These correspond to a super-Eddington efficiency lower-limit of the order of $\eta_{\rm bol} \ge 10^5$ at $d =$ \SI{1000}{\mega\parsec}, and of $\eta_{\rm bol} \ge 10^6$ at $d =$ \SI{4000}{\mega\parsec}. 
    Note that these estimates are obtained assuming that the emission is distributed isotropically. 
    Results are shown in Fig.~\ref{fig:eff}.

    These values are at least three orders of magnitude above the super-Eddington luminosity that is expected to be achievable in these systems \citep{jiang2014global, mckinney2014three, murase2016ultrafast}. 
    Therefore, it would have been impossible to detect any of the two sources in absence of strong beaming effects capable of boosting the effective luminosity by some orders of magnitude \citep{bartos2017rapid}. 
    However, the prospect of detection improves significantly when reducing the distance of the merger. 
    For instance, at \SI{100}{\mega\parsec} we could detect a source with $\eta_{\rm bol} \sim 10^3$ with only 2 hours of exposure. 

    \subsection{Post-merger Jet}

    \begin{figure*}[ht]
    \centering
    \includegraphics[width=0.8\textwidth]{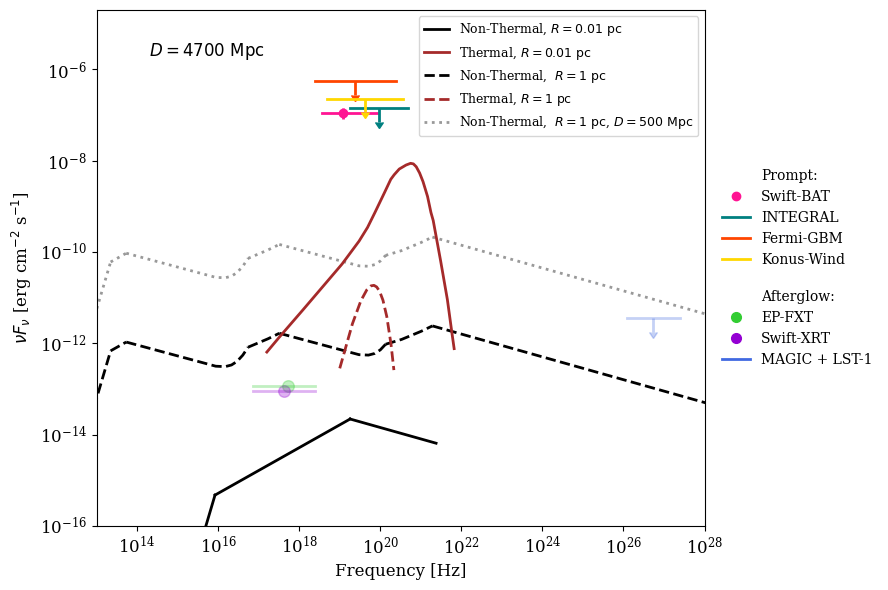}
    \caption{SED of thermal (brown) and non-thermal (black) emission from the post-merger shock in the case of a BBH system with a total mass of $M_{\text{tot}}$\,=\,120\,M$_\odot$ at \SI{4700}{\mega\parsec} from the observer. 
    The solid lines represent the case with $R=$\SI{e-2}{\parsec} and $t_\text{duration} =$ \SI{30}{\second}, and the dashed lines the case with $R=$\SI{1}{\parsec} and $t_\text{duration} =$ \SI{e5}{\second}. 
    Additionally, we plot the non-thermal emission at \SI{500}{\mega\parsec} and $R=$\SI{1}{\parsec} in light dotted black. 
    We show flux points and upper-limits obtained during the prompt-phase (i.e. a few seconds after the GW signal) with \textit{Swift}-BAT \protect\citep{2024GCN.38351....1D}, \textit{INTEGRAL} \protect\citep{2024GCN.38311....1S}, \textit{Fermi}-GBM \protect\citep{2024GCN.38316....1S}, and \textit{Konus}-Wind \protect\citep{2024GCN.38321....1R}. 
    For completeness, we also show the X-ray candidate counterpart fluxes reported by EP-FXT \protect\citep{2024GCN.38324....1P} and \textit{Swift}-XRT \protect\citep{2024GCN.38345....1W} during the afterglow (i.e. hours or days after the GW signal), together with our MAGIC+LST-1 deepest upper-limit. 
    Afterglow measurements are displayed with a lighter tone to distinguish them from the prompt observations. 
    This figure was adapted with courtesy from \protect\citet{tagawa2023observable}.
    }
    \label{fig:sed}
    \end{figure*}

    While the super-Eddington framework discussed in Sect.~\ref{sec:accretion} provides a direct insight into the energetics of the system based on a simplified parameterization, it does not model the specific physical mechanisms generating the multi-wavelength spectrum. 
    To complement this broad constraint, we now consider a highly detailed model, based on the work by \citep{tagawa2023observable}.
    
    When an accreting BH is rapidly spinning in a highly magnetized medium, a jetted emission occurs, labelled Blandford-Znajek (BZ) jet \citep{blandford1977electromagnetic, tagawa2022can}. 
    As the jet collides with gas it forms strong forward and reverse shocks, where photons are initially trapped due to scattering processes. 
    Eventually, as the shock approaches the disk surface, photons start to escape and electrons are efficiently accelerated producing non-thermal emission via synchrotron and inverse Compton scattering processes, whereas the second- and third-order inverse Compton scatterings are suppressed by the Klein-Nishina effect.

    Shock breakout emission from a BBH system is expected to be produced even before the merger, as the jet can be efficiently launched by a single accreting BH. 
    However, catching such emission without any prior information on the localization of the BBH and the nature of it is challenging -- if not impossible. 
    Hence, observations must be triggered by a GW signal. 
    In this regard, \citet{tagawa2023observable} tailored a theoretical framework that accounts for both thermal and non-thermal emission from the collision of a BZ jet emitted after merger with the un-shocked gas of an AGN disk. 
    The key element is that the post-merger jet produced by the remnant BH has to be launched with a different orientation with respect to pre-merger jets in order to interact with un-shocked gas and trigger shocks. 
    This is possible because before the merger each BH's spin tends to be aligned perpendicularly to the AGN disk, while, during the merger, the binary’s orbital angular momentum is generally randomized due to frequent binary–single interactions and inhomogeneities in the disk. 
    As a result, the remnant BH will acquire a new spin orientation, different from the pre-merger alignment, and the shock breakout emission will follow the GW signal.

    This process is regulated by two timescales: the delay time between the emission of a jet and the consequent breakout emission ($t_\text{delay}$) and the duration of emission from a breakout shell ($t_{\text{duration}}$). 
    In non-relativistic regimes, the former coincides, with the time delay between the GW and the breakout, and the latter with the duration of the bright emission itself. 
    In relativistic regimes, and when shocks are propagating towards the observer, the breakout timescale is reduced due to the beaming effect, to pair-production, and to the Klein-Nishina effect.
    These timescales are particularly sensible to the properties of the AGN disk, its density profile and the radial distance $R$ of the BBH from the central super-massive black-hole (SMBH). 
    In particular, the duration of the emission scales as $t_\text{duration} \sim R^{3}$, with the consequence that BBHs located further from the SMBH have a higher probability to be detected following a GW alert.
    This is because the usual time of pointing a ground-based telescope ($\sim$\,\SI{1}{\day}) is shorter than the expected duration of mergers at $R\ge$\SI{1}{\parsec} of $\sim$ weeks. 
    The authors refer the readers to \citet{tagawa2023observable} for further details on the assumptions and physical parameters for this model. 
 
    Fig.~\ref{fig:sed} shows an example of the predicted SED from a BBH merger with $M_{\text{tot}} = 120 M_\odot$ evaluated at the distance of GW241125 for two values of radial distance from the central SMBH (1 and \SI{0.01}{\parsec}), corresponding to two values of duration of the emission (\num{e+5} and \SI{30}{\second}). 
    For comparison, we plot the observed flux and flux upper-limits obtained with the \textit{Swift}-BAT, \textit{INTEGRAL}, \textit{Fermi}-GBM and \textit{Konus}-Wind telescopes in the (rest-frame) timescale between $T_0\,+ 6.8\,\text{s}$ and $T_0\,+ 8\,\text{s}$, with the \textit{Swift}-XRT and EP-FXT telescopes between $T_0\,+$ \SI{55}{\kilo\second} and $T_0\,+$ \SI{94}{\kilo\second}, and the upper-limits obtained with MAGIC+LST-1 at $T_0\,+$ \SI{32}{\kilo\second}. 
    Following the same nomenclature used for gamma-ray bursts, we define as "prompt" the observations performed a few seconds after the GW signal and "afterglow" the observations performed hours to days after the GW signal. 
    This distinction, however, is just conventional: the signal should not have changed drastically over the entire duration of observations if $t_{\rm duration} \sim 10^5$ s. 

    The observed $t_{\text{delay}}$ of the \textit{Swift}-BAT detection is (in the rest-frame) $11.264/(1+z) \approx$ \SI{6.47}{\second} \citep{2024GCN.38308....1D, zhang2026lvk}, while the $t_\text{duration}$ of the emission can be assumed to be equal to the time-bin in which the burst was identified, i.e. \SI{0.512}{\second}.
    These information, combined with the AGN properties constrained by \citet{zhang2026lvk}, indicate that the BBH merger was likely located deep into the AGN disk, at $R\le$\SI{e-2}{\parsec}. 
    Moreover, the \textit{Swift}-BAT flux lies significantly above the expectations from the thermal emission model in the $R=$\SI{e-2}{\parsec} scenario, suggesting that the BBH merging location was even deeper inside the disk. 
    These findings drastically reduce the possibility of a detection of VHE gamma-rays with the MAGIC+LST-1 telescopes. 
    In fact, even if the observations were conducted during the prompt emission, the flux would have been orders of magnitude dimmer than the telescopes' sensitivity. 
    Moreover, we note that this model is somehow optimistic: gamma-ray flux at GeV -- TeV energies can be severely obscured by $\gamma-\gamma$ annihilation if the Lorentz factor of the jet is not sufficiently high (i.e. $\Gamma\sim 30$; \citealt{Tagawa2023high}); also, since the peak of the observed emission lies in the X-ray regime, the VHE flux is expected to be highly suppressed by the Klein-Nishina effect, further diminishing the expected flux (see for example Figure 2 of \citealt{Tagawa2023high}).
    Additionally, even if the flux was strong enough to exceed the detection threshold, several observational challenges would remain. 
    These include complicating environmental effects such as torus obscuration, gas properties, and beaming \citep{tagawa2023observable, tagawa2024shock}.
    Nevertheless, according to this model, emission from a merger occurring at $\sim$ \SI{500}{\mega\parsec} and within $R \geq$ \SI{1}{\parsec} could, under favourable conditions, be detectable from Earth. 

\subsection{AGN association}\label{sec:agns}

\begin{deluxetable*}{cccccccc}
\tablecaption{
The three AGN candidates identified within the 95\% localization volume of GW240615.
\label{tab:AGNs}
}
\tablehead{
\colhead{Gaia DR3 Source ID} &
\colhead{R.A.} &
\colhead{Dec.} &
\colhead{$z_{\rm phot}$} &
\colhead{$D_{\rm L}$} &
\colhead{Int. prob.} &
\colhead{Flux UL} &
\colhead{Flux UL} \\
\colhead{} &
\colhead{(deg)} &
\colhead{(deg)} &
\colhead{} &
\colhead{(Mpc)} &
\colhead{} &
\colhead{$0.15<E<0.6$ TeV} &
\colhead{$E>0.6$ TeV}
}
\startdata
388920283769824128 & 7.18 & 45.91 & 0.27(7) & 1434(413) & 0.37 & $1.8\times10^{-11}$ & $1.1\times10^{-12}$ \\
389394242001818624 & 7.01 & 46.77 & 0.34(10) & 1840(623) & 0.89 & $2.4\times10^{-11}$ & $2.3\times10^{-12}$ \\
389182689092831872 & 8.96 & 46.52 & 0.33(13) & 1818(791) & 0.83 & $3.7\times10^{-11}$ & $1.5\times10^{-12}$ \\
\enddata

\tablenotetext{}{
IDs and coordinates are from Gaia DR3 \citep{gaia_dr3_egal} and unWISE
\citep{Schlafly_2019_unWISE}. Photometric redshifts ($z_{\rm phot}$)
are from the Quaia catalog. Luminosity distances ($D_{\rm L}$) were
computed assuming a flat $\Lambda$CDM cosmology with
$H_0 = 67.4\,{\rm km\,s^{-1}\,Mpc^{-1}}$ and
$\Omega_{\rm m}=0.315$ \citep{refId0_planck_cosmo}.
We also report the integrated 2D probability of association with the GW
event and the flux upper limits for each source.
}
\end{deluxetable*}

    The identification of a host galaxy or AGN counterpart to a GW event relies on the completeness of the candidate catalogues available. For both events several catalogues were investigated: GLADE+ (optical/near-IR; \citealt{10.1093/mnras/stac1443_glade+}), Quaia (optical/near-IR; \citealt{Storey-Fisher_2024_quaia}), MILLIQUAS v8 (optical; \citealt{2023OJAp....6E..49F_milliquas}), and MORX v2 (radio/X-ray association catalog; \citealt{2016PASA...33...52F_morx}). 
    Note that the redshift present in most of those big catalogues is photometric, which implies a factor $~100$ larger than spectrometric redshifts, affecting the association reliability.
    We also searched the \textit{Fermi}-LAT 4th source catalogue \citep{2020ApJS..247...33A_fermi4thfgl}, finding no high-energy sources for any of the regions.
    
    Considering a fiducial AGN number density $\rho_{\rm agn}$ =  $10^{-4.75}\mathrm{M\href{}{}pc}^{-3}$ \citep{greene2009erratum,bartos2017gravitational}, a total of $\sim30$ AGN are expected in the GW240615 95\% volume. 
    Despite the precise 2D localization of GW241125, its distance uncertainty leads to a comparable expectation of $\sim$10 AGNs within the search volume.

    Galactic extinction and stellar crowding at low galactic latitudes (\SI{-16.9}{\degree} and \SI{12.1}{\degree} respectively) limit the completeness of extragalactic catalogues for GW240615 and GW241125. 
    In the case of the first one, the \SI{95}{\percent} localization volume yielded 3 AGNs and $\sim$2000 field galaxies. 
    This is $\sim$\SI{10}{\percent} of the expected density of both, confirming a deficit in catalogue completeness. 
    In the case of GW241125, given the proximity to galactic plane and larger distance, no AGNs or galaxies are found in the \SI{95}{\percent} volume.

    Despite the catalogue incompleteness, the crossmatch for GW240615 identified 3 AGNs, all from Quaia catalogue. Details of these 3 sources in Table~\ref{tab:AGNs} (also show in Fig.~\ref{fig:maps-flux}). None of the identified candidates spatially coincide with any localized region of elevated significance.
    Recent estimates based on comparisons to the observed BH masses, spins, and merger rate, suggest that BBH mergers happening inside of AGN disk can contribute to $25-80\%$ of all GWs detected by LVK~\citep{ford2022binary}. 
    Given the observed deficit in catalog completeness near the galactic plane and the lack of candidates for GW241125, and the uncertainty in the redshift, we cannot definitively claim nor exclude that either event originated within an AGN disk. 

\section{Conclusions}\label{Sec:8}

    We presented VHE follow-up of two GW candidates, GW240615 and GW241125. 
    The sources were observed with the MAGIC and CTAO LST-1 telescopes both in standalone and stereo joint configurations, operating in a standard wobble mode, which allowed the longest on-source search of EM counterpart candidates in the VHE band ever obtained with IACT systems. 
    GW240615 was the best localized event of all LVK runs, with a region of uncertainty of only \SI{5}{\square\deg}, smaller than the FoV of the MAGIC and LST-1 telescopes. 
    GW241125 was the only event during O4 with a putative EM candidate detected with the \textit{Swift}-BAT telescope during the prompt emission and with the \textit{Swift}-XRT and EP-FXT during the afterglow.  
    No detection was achieved in the GeV -- TeV band for any of the two sources. 
    In the case of GW240615, we used an existing method to combine the observation significance sky-map and the GW probability area to achieve a single limit. 
    We obtained \SI{5.17e-11}{\per\square\centi\meter\per\second} in the 0.15 -- \SI{0.6}{\tera\electronvolt} energy range and \SI{4.47e-12}{\per\square\centi\meter\per\second} above \SI{0.6}{\tera\electronvolt}. 
    In the case of GW241125, we performed a tailored analysis for sources with suboptimal weather conditions and constrained the VHE of the putative \textit{Swift}-BAT source, achieving an upper-limit of \SI{2.3e-11}{\per\square\centi\meter\per\second} above \SI{0.5}{\tera\electronvolt}. 
    We discussed two state-of-the-art physical mechanisms that could produce an observable VHE gamma-ray emission from a BBH merger embedded in the disk of an AGN: 
    \begin{enumerate}
    
        \item The super-Eddington accretion onto a BBH system may produce observable EM signatures in proximity of the merger, however no considerations are made regarding the time delay with respect to the GW emission and the particle acceleration efficiency which may give rise to a VHE emission.
        Assuming a conversion factor of $\epsilon_\gamma = 0.1$, defined as the fraction of photons that are efficiently up-scattered up to VHE energies, we showed that, to be detected by our telescopes, a BBH merger similar to GW240615 and GW241125 should accrete with a super-Eddington efficiency of $\eta_{\rm bol} \ge 10^6$, which is several orders of magnitude above the known values of other super-Eddington accreting systems. 

        \item The shock breakout post-merger produced by the collision of a Blandford-Znajek jet with the dense gaseous disk of the AGN may give rise to thermal and non-thermal emission, on timescales that depend on the AGN properties, the localization of the BBH in the system and the distance of the event. 
        The VHE emission is expected from inverse Compton scattering mechanism and it is highly dependent on the distance $R$ of the merger from the central SMBH. 
        A VHE signature from BBH events similar to GW240615 and GW241125 may be strong enough to be detected by MAGIC+LST-1 if $R \ge 1$, other than being more feasible for a follow-up by a ground-based telescope.  
        However, we showed how, in the case of GW241125, the detection of a counterpart by \textit{Swift}-BAT a few seconds after the GW rules out this possibility and indicates that the system was probably deeper inside the disk.  
    \end{enumerate}
    
    None of the mechanisms discussed above appears capable of producing a detectable VHE signal from GW240615 or GW241125.
    In addition, several challenges arise that are not directly related to the physical mechanisms behind such events: observing conditions, the time delay with respect to the GW signal, day–night constraints of our telescopes, and uncertainties regarding the true origin of the GWs.
    Nevertheless, our constraints suggest that a nearby event ($\leq$ \SI{500}{\mega\parsec}) could be detectable in the future, particularly given the expected increase in sensitivity with the full CTAO array \citep{cta2018science}.

    This work establishes a robust precedent for the VHE follow-up of GWs. 
    Beyond the specific limits set on GW240615 and GW241125, the methodological frameworks developed here should serve as a base for future works treating sources with uncertain localization and sources observed in suboptimal weather conditions. 
    Furthermore, our exploration of emission models extends the existing literature of GW counterparts observed with IACTs, preparing the necessary grounds for the upcoming CTAO.

\begin{acknowledgments}
A. Simongini gratefully acknowledges H. Tagawa for valuable discussions on theoretical aspects.
\\ \\ 
\textit{MAGIC collaboration acknowledgements:} 
We would like to thank the Instituto de Astrof\'{\i}sica de Canarias for the excellent working conditions at the Observatorio del Roque de los Muchachos in La Palma. The financial support of the German BMFTR, MPG and HGF; the Italian INFN and INAF; the Swiss National Fund SNF; the grants PID2022-136828NB-C41, PID2022-137810NB-C22, PID2022-138172NB-C41, PID2022-138172NB-C42, PID2022-138172NB-C43, PID2022-139117NB-C41, PID2022-139117NB-C42, PID2022-139117NB-C43, PID2022-139117NB-C44, CNS2023-144504 funded by the Spanish MCIN/AEI/ 10.13039/501100011033 and "ERDF A way of making Europe"; the Indian Department of Atomic Energy; the Japanese ICRR, the University of Tokyo, JSPS, and MEXT; the Bulgarian Ministry of Education and Science, National RI Roadmap Project D01-135/16.07.2025 and the Academy of Finland grant nr. 320045 is gratefully acknowledged. This work has also been supported by Centros de Excelencia ``Severo Ochoa'' y Unidades ``Mar\'{\i}a de Maeztu'' program of the Spanish MCIN/AEI/ 10.13039/501100011033 (CEX2019-000918-M, CEX2021-001131-S, CEX2024-001441-S), by AST22\_00001\_9 with funding from NextGenerationEU funds and by the CERCA institution and grants 2021SGR00426, 2021SGR00607 and 2021SGR00773 of the Generalitat de Catalunya; by the Croatian Science Foundation (HrZZ) Project IP-2022-10-4595 and by the University of Rijeka Project uniri-mzi-25-3 funded by the European Union - NextGenerationEU; by the Deutsche Forschungsgemeinschaft (SFB1491) and by the Lamarr-Institute for Machine Learning and Artificial Intelligence; by the Polish Ministry of Science and Higher Education grant No. 2025/WK/04; by the European Union (ERC/MicroStars, 101076533 and INFRA/ACME, 101131928); and by the Brazilian MCTIC, the CNPq Productivity Grant 309053/2022-6 and FAPERJ Grants E-26/200.532/2023 and E-26/211.342/2021.
\\ \\ 
\textit{LST collaboration acknowledgements:} We gratefully acknowledge financial support from the following agencies and organisations:
Conselho Nacional de Desenvolvimento Cient\'{\i}fico e Tecnol\'{o}gico (CNPq) Grant 309053/2022-6 and Funda\c{c}\~{a}o de Amparo \`{a} Pesquisa do Estado do Rio de Janeiro (FAPERJ) Grants E-26/200.532/2023 and E-26/211.342/2021, Funda\c{c}\~{a}o de Amparo \`{a} Pesquisa do Estado de S\~{a}o Paulo (FAPESP), Funda\c{c}\~{a}o de Apoio \`{a} Ci\^encia, Tecnologia e Inova\c{c}\~{a}o do Paran\'a - Funda\c{c}\~{a}o Arauc\'aria, Ministry of Science, Technology, Innovations and Communications (MCTIC), Brasil;
the Bulgarian Ministry of Education and Science, National RI Roadmap Project D01-135/16.07.2025, Bulgaria;
Croatian Science Foundation (HrZZ) Project IP-2022-10-4595; University of Rijeka Project uniri-mzi-25-3 funded by the European Union - NextGenerationEU; the following Croatian Institutions: Faculty of Physics, University of Rijeka; University Josip Juraj Strossmayer of Osijek; Faculty of Electrical Engineering, Mechanical Engineering and Naval Architecture (FESB), University of Split, Croatia;
Ministry of Education, Youth and Sports, MEYS  LM2023047, EU/MEYS CZ.02.1.01/0.0/0.0/16\_013/0001403, \\ CZ.02.1.01/0.0/0.0/18\_046/0016007, \\ CZ.02.1.01/0.0/0.0/16\_019/0000754, \\ CZ.02.01.01/00/22\_008/0004632 and \\ CZ.02.01.01/00/23\_015/0008197 Czech Republic;
CNRS-IN2P3, the French Programme d’investissements d’avenir and the Enigmass Labex, 
This work has been done thanks to the facilities offered by the Univ. Savoie Mont Blanc - CNRS/IN2P3 MUST computing center, France;
Max Planck Society, German Bundesministerium f{\"u}r Forschung, Technologie und Raumfahrt (Verbundforschung / ErUM), the Deutsche Forschungsgemeinschaft (SFB 1491) and the Lamarr-Institute for Machine Learning and Artificial Intelligence, Germany;
Istituto Nazionale di Astrofisica (INAF), Istituto Nazionale di Fisica Nucleare (INFN), Italian Ministry for University and Research (MUR), and the financial support from the European Union -- Next Generation EU under the project IR0000012 - CTA+ (CUP C53C22000430006), announcement N.3264 on 28/12/2021: ``Rafforzamento e creazione di IR nell’ambito del Piano Nazionale di Ripresa e Resilienza (PNRR)'';
ICRR, University of Tokyo, JSPS, MEXT, Japan;
JST SPRING - JPMJSP2108;
Narodowe Centrum Nauki, grant number 2023/50/A/ST9/00254, Poland;
The Spanish groups acknowledge the Spanish Ministry of Science and Innovation and the Spanish Research State Agency (AEI) through the government budget lines
PGE2022/28.06.000X.711.04,
28.06.000X.411.01 and 28.06.000X.711.04 of PGE 2023, 2024 and 2025,
and grants PID2019-104114RB-C31,  PID2019-107847RB-C44, PID2019-105510GB-C31, PID2019-104114RB-C33, PID2019-107847RB-C43, PID2019-107847RB-C42, PID2019-107988GB-C22, PID2021-124581OB-I00, PID2021-125331NB-I00, PID2022-136828NB-C41, PID2022-137810NB-C22, PID2022-138172NB-C41, PID2022-138172NB-C42, PID2022-138172NB-C43, PID2022-139117NB-C41, PID2022-139117NB-C42, PID2022-139117NB-C43, PID2022-139117NB-C44, PID2022-136828NB-C42, PID2024-155316NB-I00, PDC2023-145839-I00 funded by the Spanish MCIN/AEI/10.13039/501100011033 and by ERDF/EU and NextGenerationEU PRTR; CSIC PIE 202350E189; the "Centro de Excelencia Severo Ochoa" program through grants no. CEX2020-001007-S, CEX2021-001131-S, CEX2024-001441-S; the "Unidad de Excelencia Mar\'ia de Maeztu" program through grants no. CEX2019-000918-M, CEX2020-001058-M, CEX2024-001451-M; the "Ram\'on y Cajal" program through grants RYC2021-032991-I  funded by MICIN/AEI/10.13039/501100011033 and the European Union “NextGenerationEU”/PRTR and RYC2020-028639-I; the "Juan de la Cierva-Incorporaci\'on" program through grant no. IJC2019-040315-I and "Juan de la Cierva-formaci\'on"' through grant JDC2022-049705-I; the “Viera y Clavijo” postdoctoral program of Universidad de La Laguna, funded by the Agencia Canaria de Investigaci\'on, Innovaci\'on y Sociedad de la Informaci\'on. They also acknowledge the "Atracci\'on de Talento" program of Comunidad de Madrid through grant no. 2019-T2/TIC-12900; “MAD4SPACE: Desarrollo de tecnolog\'ias habilitadoras para estudios del espacio en la Comunidad de Madrid" (TEC-2024/TEC-182) project, Doctorado Industrial (IND2024/TIC34250) and Ayudas para la contrataci\'on de personal investigador predoctoral en formación (PIPF-2023/TEC-29694) funded by Comunidad de Madrid; the La Caixa Banking Foundation, grant no. LCF/BQ/PI21/11830030; Junta de Andaluc\'ia under Plan Complementario de I+D+I (Ref. AST22\_0001) and Plan Andaluz de Investigaci\'on, Desarrollo e Innovaci\'on as research group FQM-322; Project ref. AST22\_00001\_9 with funding from NextGenerationEU funds; the “Ministerio de Ciencia, Innovaci\'on y Universidades”  and its “Plan de Recuperaci\'on, Transformaci\'on y Resiliencia”; “Consejer\'ia de Universidad, Investigaci\'on e Innovaci\'on” of the regional government of Andaluc\'ia and “Consejo Superior de Investigaciones Cient\'ificas”, Grant CNS2023-144504 funded by MICIU/AEI/10.13039/501100011033 and by the European Union NextGenerationEU/PRTR,  the European Union's Recovery and Resilience Facility-Next Generation, in the framework of the General Invitation of the Spanish Government's public business entity Red.es to participate in talent attraction and retention programmes within Investment 4 of Component 19 of the Recovery, Transformation and Resilience Plan; Junta de Andaluc\'{\i}a under Plan Complementario de I+D+I (Ref. AST22\_00001), Plan Andaluz de Investigaci\'on, Desarrollo e Innovación (Ref. FQM-322). ``Programa Operativo de Crecimiento Inteligente" FEDER 2014-2020 (Ref.~ESFRI-2017-IAC-12), Ministerio de Ciencia e Innovaci\'on, 15\% co-financed by Consejer\'ia de Econom\'ia, Industria, Comercio y Conocimiento del Gobierno de Canarias; the "CERCA" program and the grants 2021SGR00426 and 2021SGR00679, all funded by the Generalitat de Catalunya; and the European Union's NextGenerationEU (PRTR-C17.I1). This work is funded/Co-funded by the European Union (ERC, MicroStars, 101076533). This research used the computing and storage resources provided by the Port d'Informaci\'o Cient\'ifica (PIC) data center.
State Secretariat for Education, Research and Innovation (SERI) and Swiss National Science Foundation (SNSF), Switzerland;
The research leading to these results has received funding from the European Union's Seventh Framework Programme (FP7/2007-2013) under grant agreements No~262053 and No~317446;
This project is receiving funding from the European Union's Horizon 2020 research and innovation programs under agreement No~676134;
ESCAPE - The European Science Cluster of Astronomy \& Particle Physics ESFRI Research Infrastructures has received funding from the European Union’s Horizon 2020 research and innovation programme under Grant Agreement no. 824064.
\\ \\ 
AS, P. Gliwny, J. Sitarek, A. Stamerra   acknowledge partial financial support by mobility "Canaletto" Program Italy-Poland 2025-2026  (MAECI: PL25MO12 and NAWA:BPN/BIT/2024/1/00037).
\\ \\ 
A. Stamerra acknowledges financial support from INAF through the “Ricerca Fondamentale 2024” on the GO/GTO program titled ‘Discovering the New TeV Frontier of Gravitational Wave Counterparts through Follow-up Observations with the MAGIC Telescopes'.
\end{acknowledgments}

\begin{contribution}

    A. Simongini: project coordination, LST-1 and MAGIC data analysis, theoretical modelling, discussion and interpretation, paper drafting and editing. 
    J. Jiménez Quiles: project coordination, main LST-1 and MAGIC data analysis, statistical methods, paper drafting and editing.
    M. Seglar Arroyo: project coordination, PI of the MAGIC+LST-1 proposal, AGN association analysis. 
    M. Pecimotika: data analysis, paper drafting and editing. 
    A. Moralejo: data analysis supervision and discussion
    A. Stamerra: PI of the MAGIC+LST-1 proposal, discussion. 
    A. Carosi: discussion.
    S. Menon: fast offline analysis. 
    
    All authors above have participated in the paper discussion and edition.
    The rest of the authors have contributed in one or several of the following ways: design, construction, maintenance and operation of the instrument(s) used to acquire the data; preparation and/or evaluation of the observation proposals; data acquisition, processing, calibration and/or reduction; production of analysis tools and/or related Monte Carlo simulations; discussion and approval of the contents of the draft.

\end{contribution}

\facilities{MAGIC, CTAO LST-1}

\software{\texttt{gammapy v2.0} \citep{acero_2025_17814297_gammapy2.0}, \texttt{cta-lstchain v0.10.7} \citep{lstchain-Zenodo_2024}, \texttt{MARS} \citep{2013ICRC...33.2937Z}, \texttt{magic-cta-pipe} \citep{abe2023performance}, and packages therein. 
          }

\appendix
\section{Addressing the NSB conditions and varying energy thresholds of observations}\label{appendix:nsb-settings-e-threshold}

\subsection{GW240615}

\begin{figure}[H]
        \includegraphics[width=1\columnwidth]{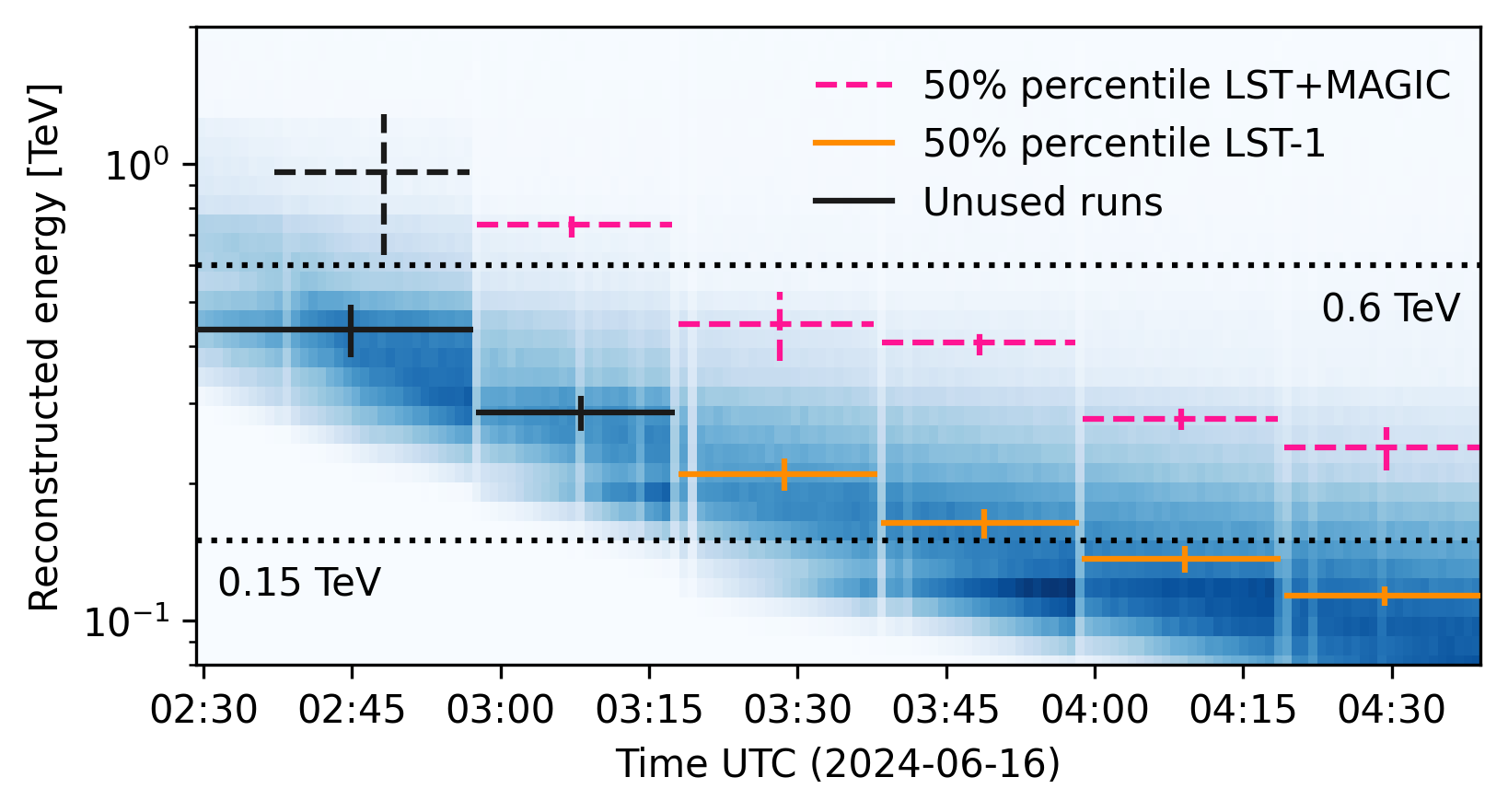}
        \caption{Run-wise median of the reconstructed event energy for GW240615, used to estimate energy threshold in the case of GW240615. The energy threshold is set in order to have enough background events to create a reliable background model. Horizontal lines separate the energy regions where the \SI{50}{\percent} of events are contained. Error bars represent the standard deviation of the percentile values, estimated over short time slices within each run. In the background histogram the total LST-1 distributions are shown in linear scale. The gaps results from changes of pointing and in some cases data removed because of the car flashes.}
        \label{fig:energy-th-GW240615}
    \end{figure}

    Observations of GW240615 were performed at relatively high zenith angles ranging \SI{65}{\deg} to \SI{45}{\deg} and after the Moon set on the horizon.     
    In general, at large zenith angles the energy threshold can increase substantially, reaching values close to \SI{1}{\tera\electronvolt} in some cases \citep{aleksic2016major, 2025EPJWC.31905006A}.
    Additionally, LST-1 alone achieves an energy threshold about 2.5 times lower than that of MAGIC \citep{Lopez-Coto2021performanceLST, abe2023performance}, enabling access to gamma rays at much lower energies.
    When combining data in stereo mode, requiring simultaneous observations with at least two telescopes, the resulting performance reflects the strengths of both configurations: MAGIC+LST-1 observations offer improved sensitivity at higher energies, while LST-1 alone provides better coverage in the low-energy range. We checked the totality of events distribution, composed mostly from hadronic background gamma-like events, as we need enough of these type to create a reliable background model (see Fig.~\ref{fig:energy-th-GW240615}). In order to do that we find the median energy of the distributions run-wise, so we know where the bulk of the background statistics lies.

    This complementarity makes it possible to tailor analyses to different energy domains and fully exploit the capabilities of the array.
    In particular, to get the best from our data, we performed two separate analyses: for the low-energy range between \SI{0.15}{\tera\electronvolt}\,$< E_\gamma <$\,\SI{0.6}{\tera\electronvolt} we analysed LST-1 data alone, while for the high-energy range above $E_\gamma >$\,\SI{0.6}{\tera\electronvolt} we analysed MAGIC+LST-1 stereo data.

   \begin{figure}
    \includegraphics[width=1\columnwidth]{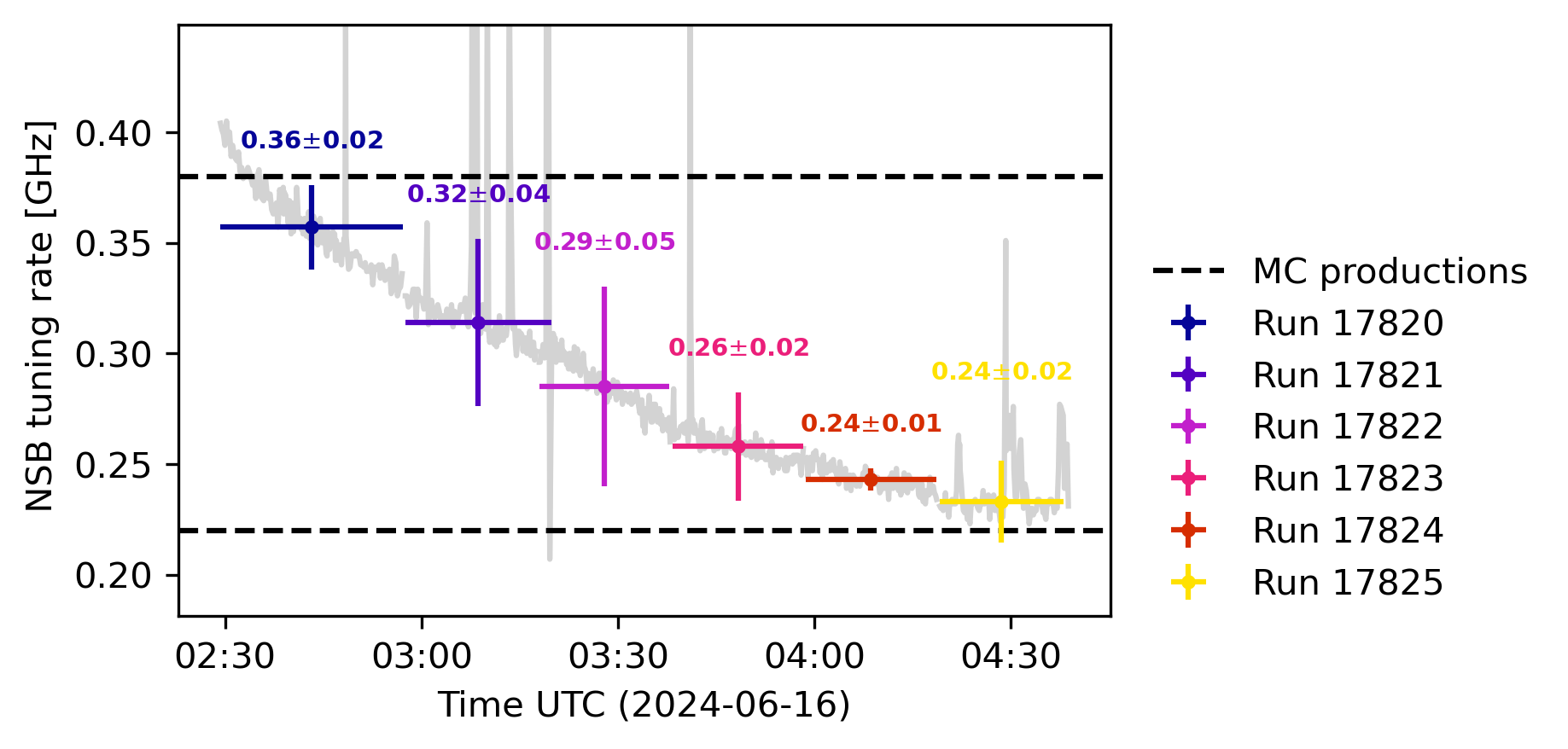}
    \caption{Variation of NSB level for images and waveforms for GW240615 observations with LST-1. The MC available productions are showed as dotted lines, and the sub-run wise values are plotted in gray in the background. The spikes that can be seen are produced by car-flashes, satellites or other light entering LST-1 camera.}
        \label{fig:nsb-tuning-GW240615}
    \end{figure}
    
   \begin{figure*}
        \includegraphics[width=1\linewidth]{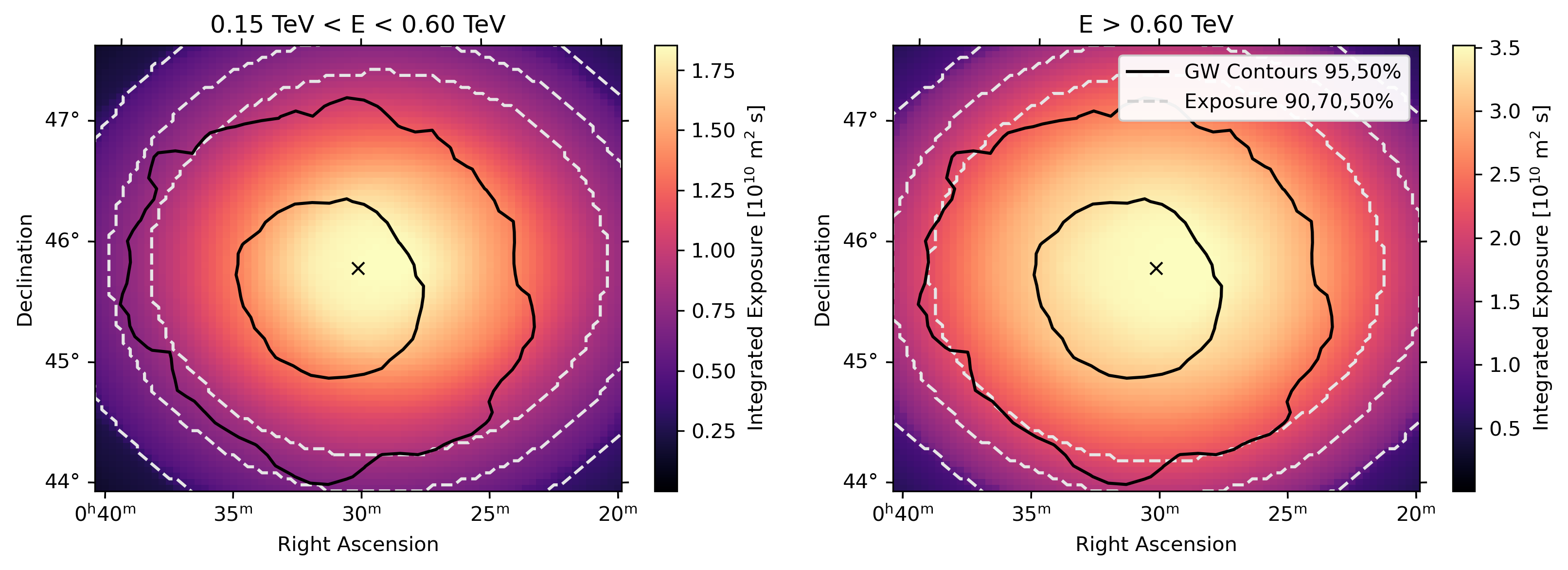}
        \caption{Integrated exposure of observations of GW241125. The exposure is computed for every energy bin of the analysis but here we integrated over all energy bins to get a unique exposure value of each bin. The contours of the \SI{50}{\percent} and \SI{95}{\percent} of the GW probability are shown as black solid lines. The 10th, 30th and 50th percentiles of the exposure are superposed in white dotted lines for each case.}
        \label{fig:maps-exposure}
    \end{figure*}

    Moreover, the observations are affected by a significantly varying NSB level.
    As shown in Fig.~\ref{fig:nsb-tuning-GW240615}, the NSB level decreased by approximately \SI{45}{\percent} over the course of two hours of observations, relative to the beginning of the run.
    These values are unusually high compared to typical extragalactic observations due to moonlight from the Moon being just below the horizon.
    
    To account for these strong variations, Poisson noise is introduced into the simulated data. 
    This can be either added at the waveform level (i.e. for each pixel signal, as implemented in \texttt{lstchain}) or at the image level (i.e. adding general noise to every pixel after the integration of signal, as implemented in \texttt{magic-cta-pipe}). 
    We selected the most appropriate NSB tuning settings on a run-wise basis from the standard set of simulated values. 
    For image-level noise, we considered NSB values of 2.0 and 2.5, expressed in units of the average number of photoelectrons per pixel. The NSB contribution was added as the standard deviation (square root) in the individual pixels. 
    For waveform-level noise, the available levels used were \SI{0.22}{\giga\hertz} and \SI{0.38}{\giga\hertz}.

\subsection{GW241125}
    \begin{figure}
        \includegraphics[width=1\columnwidth]{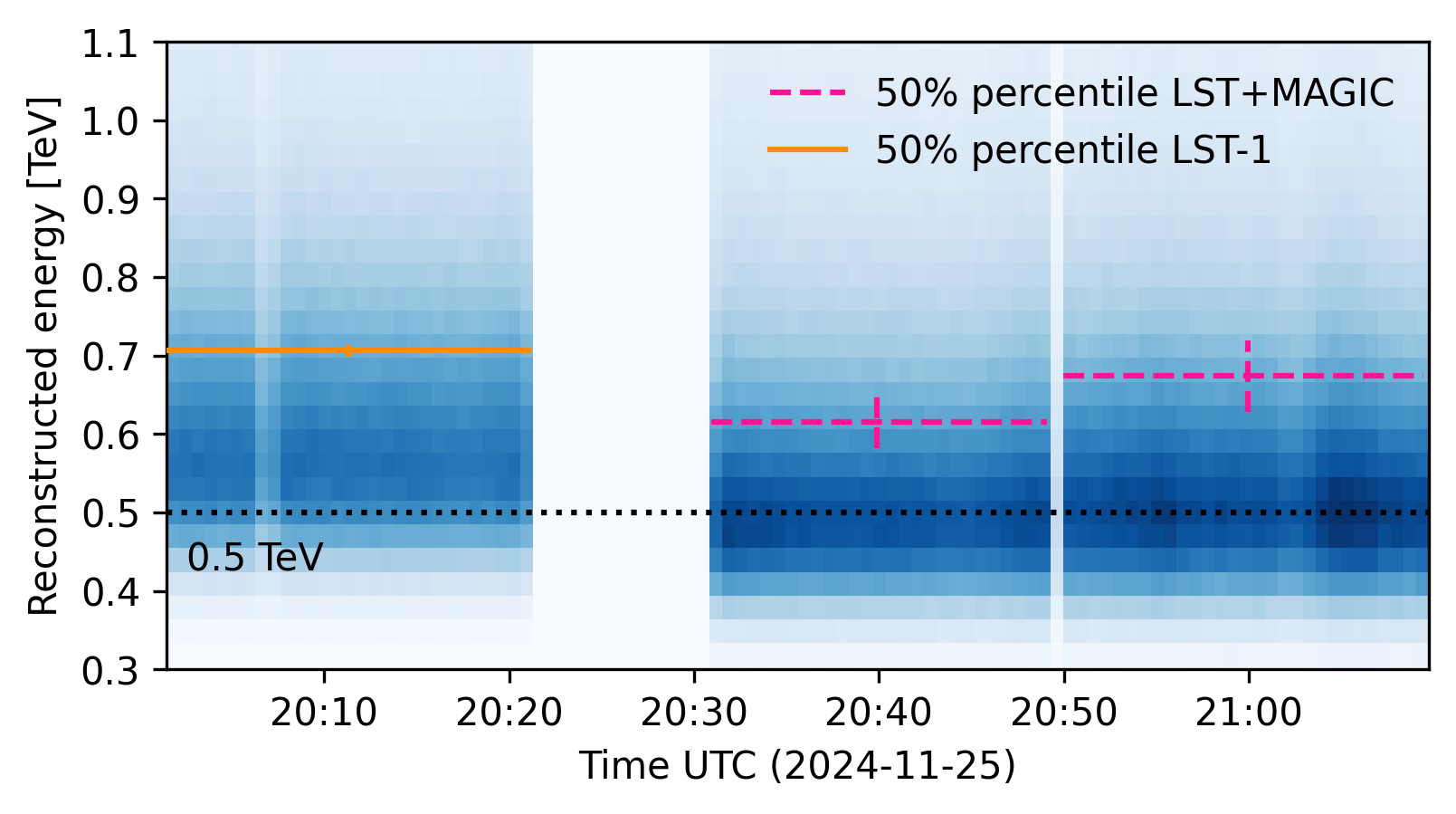}
        \caption{The run-wise median of the reconstructed event energy for GW241125 is represented in order to estimate energy threshold evolution. In the background, the total LST-1 event distributions are shown in linear scale. In orange, the LST-only runs used, in pink the LST+MAGIC runs used and in black the unused runs. The 50\% percentile might not match with the peak of distributions in the background due to long tails towards larger energies.}
        \label{fig:energy-th-GW241125}
    \end{figure}

    For GW241125, we observed the source under suboptimal atmospheric conditions, including a cloud layer at $\sim$\SI{6}{\kilo\meter} a.g.l.
    To illustrate the sky coverage achieved in these observations, we provide in Fig.~\ref{fig:maps-exposure} the exposure map of GW241125. 
    The map is obtained by summing the exposures across all reconstructed energy bins, and relatives 10th, 30th and 50th percentile thresholds are shown to highlight the regions with sufficient observational depth. This allows us to verify that the \SI{95}{\percent} localization region is fully observed.
    We represented the reconstructed energy distributions over the observations in Fig.~\ref{fig:energy-th-GW241125}. 
    The energy threshold seems to drop both for LST alone as for LST+MAGIC data, the general trend is given by the decrease of zenith angle. 
    Given the exceptional suboptimal conditions the energy threshold is highly increased, and events at low energies contribute little due to their poor reconstruction quality. 
    In order to go into a safe energy region, we analyse the data with reconstructed energy $E>$\SI{0.5}{\tera\electronvolt}.

    \section{Modifications to the standard \texttt{gammapy} analysis}\label{appendix:gammapyModifications}

    The standard \texttt{gammapy v2.0} pipeline used for this analysis required two simple modifications in order to obtain meaningful upper limit sky-maps, as some needed options are not implemented in this version. 

    The class \texttt{ExcessMapEstimator} that is used to compute the sky-maps can yield negative flux values, and therefore, in some cases, the flux upper limit can be negative as well, leading to "non-physical" results. To ensure non-negative upper limits, in the pixels with negative excess, we set the number of observed counts equal to the background expectation, which will produce non-negative conservative upper limits. The limits are conservative in the sense that they will not "profit" from a deficit of events relative to the background expectation, lest it might be due to systematics, rather than just a background fluctuation.

    The \texttt{ExcessMapEstimator} computes flux sky-maps by integrating counts within a correlation radius (see Sec.\ref{Subsec:sources-uncertain-loc}), but we are interested in the {\it total} flux from a point-like source instead. To obtain the needed conversion factor, a strong point-like source is simulated, using the same IRFs and analysis configuration as for the actual observation. The ratio between the injected flux and the average flux recovered within the correlation radius is the multiplicative factor needed to obtain the final flux upper limits on the bottom panels of Fig. \ref{fig:maps-flux}.

    \section{Results: Significance of observations}\label{appendix:SignificanceMaps}
    \subsection{Significance of GW240615}
    
    First we will discuss the significance of detection of the source GW240615. 
    As the source location is not known, it is necessary to produce significance sky-maps, which were generated following the standard method described in first part of Sect.~\ref{Subsec:sources-uncertain-loc}, with the resulting maps shown in Fig.~\ref{fig:maps-flux}. 
    No excess regions above the pre-trial $5\sigma$ threshold were detected.

     \begin{figure}[H]
        \includegraphics[width=1\columnwidth]{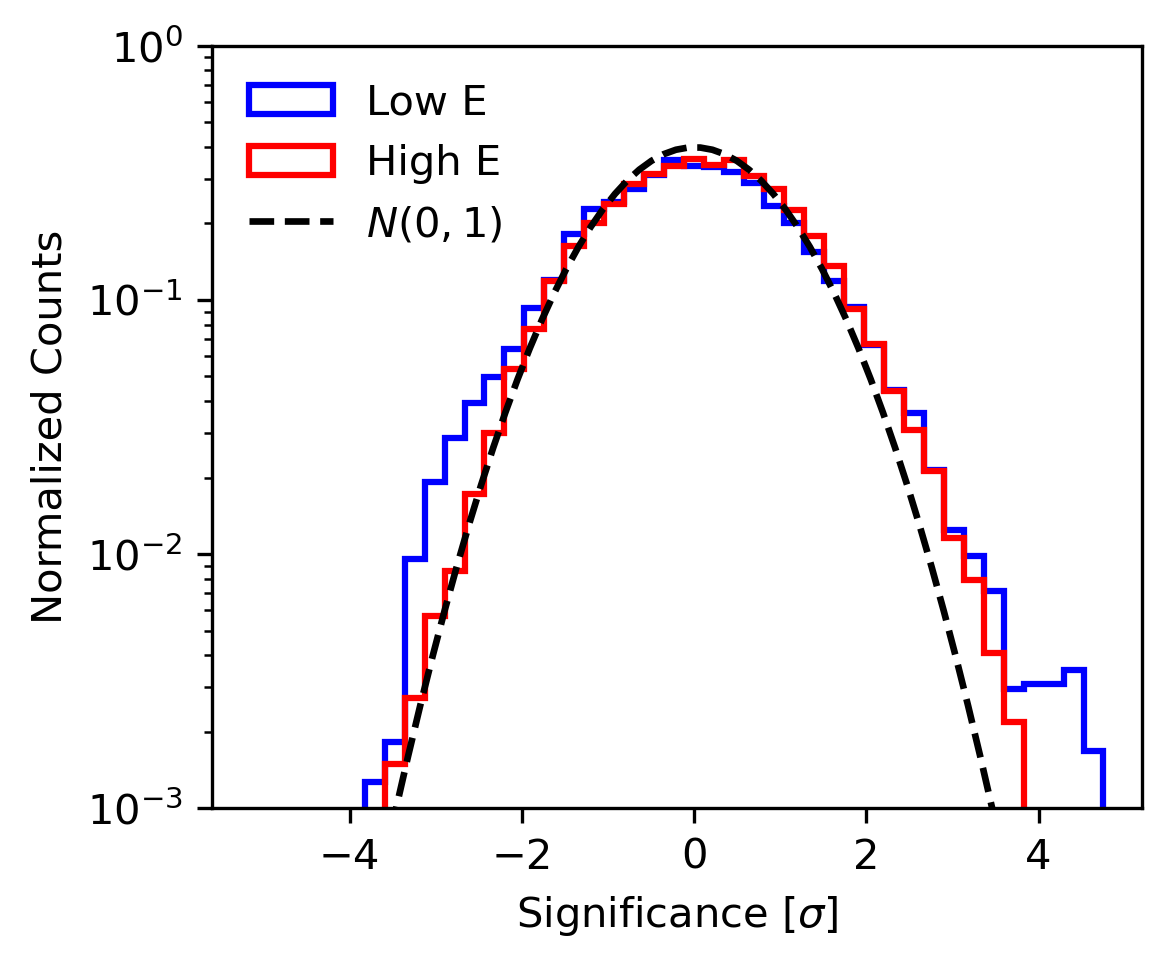}
        \caption{Projected distribution of significances for every pixel in the 95\% interest region of the GW of the sky-map for GW240615, in different energy ranges. The ideal Gaussian ($\mu=0$, $\sigma=1$) is shown as a comparison.}
        \label{fig:significance-2d}
    \end{figure}

    We validated the background reconstruction using the projected distribution of pixel significances (Fig.~\ref{fig:significance-2d}). 
    This validation is done expecting no source in the field of view. 
    In order to validate this we inspected all the background models in search of any inhomogeneity that can correspond to a source. 
    We performed a Gaussian fit over the sky-maps and we get for the low-energies case $\mu=-0.07$ and $\sigma=1.24$ while for high-energies $\mu=0.08$ and $\sigma=1.09$. 
    In both energy ranges, the Gaussian fit to the distribution closely follows a normal distribution with mean $\mu = 0$ and standard deviation $\sigma = 1$, as expected for a sky-map dominated by Poissonian noise fluctuations.

\subsection{Significance of GW241125}

     \begin{figure*}
        \includegraphics[width=1\linewidth]{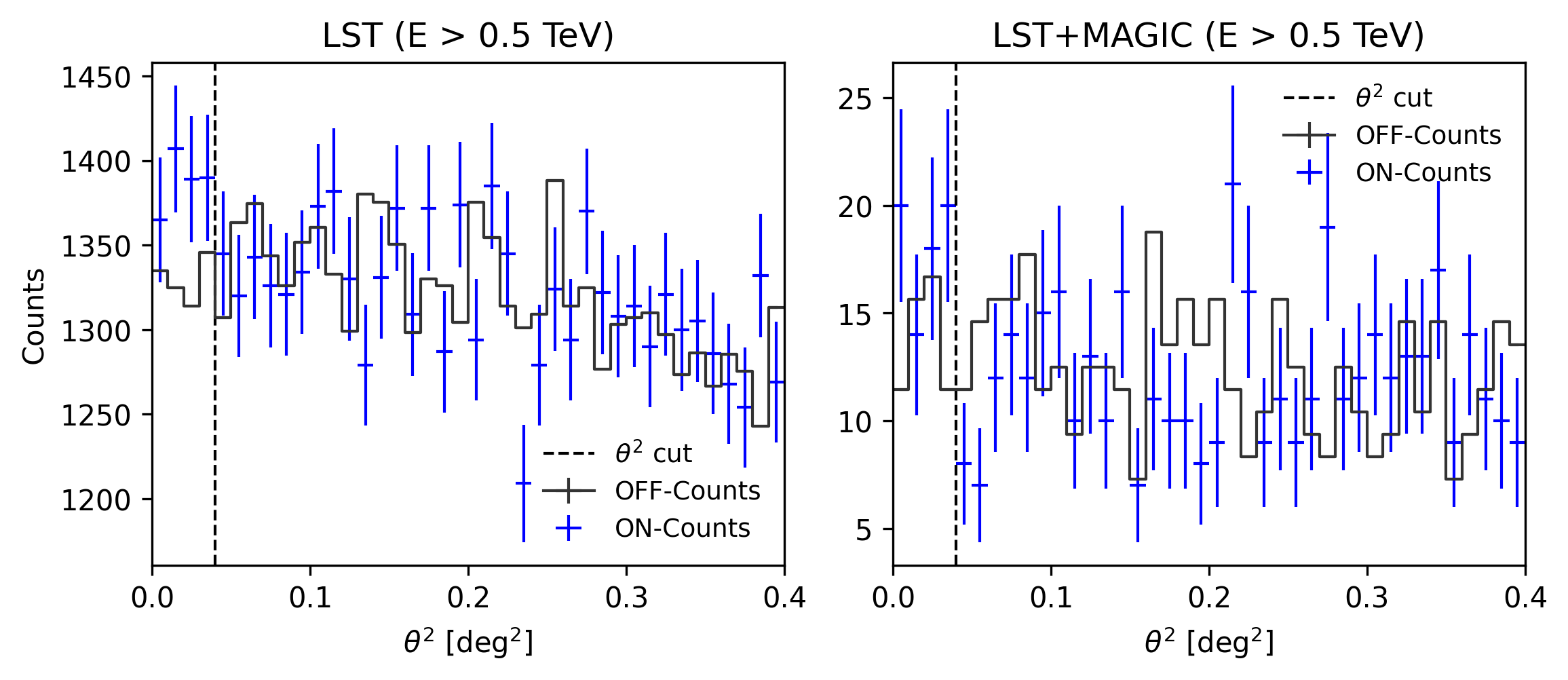}
        \caption{Distribution of squared angular distances of reconstructed events to the source position for GW241125. We show separately the LST-only run and the LST+MAGIC runs (where all events are stacked), and shown both for reconstructed energies $E>$ \SI{0.5}{\tera\electronvolt}. The obtained significances for both analysis are \num{2.2}$\sigma$ for LST-only and \num{1.47}$\sigma$ for LST+MAGIC data.}
        \label{fig:theta-plot}
    \end{figure*} 
    
    We computed the significance of the source GW241125 using Li\&Ma significance calculation \citealp{1983ApJ...272..317L_li_and_ma_significance} based on the counts in the ON and OFF regions. We binned the data in quadratic angular ($\theta^2$) bins so the area in the sky is constant for all bins. Then we set a $\theta$ cut of \SI{0.2}{\deg} to compute the number of events in both regions. Then we set 3 OFF regions symmetrically around the pointing position so the normalization factor $\alpha=1/3$. Using Li\&Ma significance formula we obtained a significance of \num{2.2}$\sigma$ for the LST data and \num{1.47}$\sigma$ for LST+MAGIC observations (see Fig.\ref{fig:theta-plot}).

\bibliography{Library_of_Alexandria}{}
\bibliographystyle{aasjournalv7}

\end{document}